\documentclass[prl,
aps,
amsmath,
amssymb,
twocolumn,
footinbib,
superscriptaddress,
longbibliography
]{revtex4-2}
\usepackage{ragged2e}    \usepackage{marginnote}   
\usepackage{cases}
\usepackage[makeroom]{cancel}
\usepackage[export]{adjustbox}
\usepackage{hyperref} 
\usepackage{caption,subcaption}
\usepackage{enumerate}
\usepackage{xr-hyper}
\usepackage{xcolor}
\begin{document}
\title{Coherently enhanced radiation friction in laser-plasma collisions}
\author{E.~G. Gelfer}\email{egelfer@gmail.com}
\affiliation{ELI Beamlines facility, The Extreme Light Infrastructure ERIC,
Dolni Brezany 252 41, Czech Republic}
\author{A.~M.~Fedotov}
\affiliation{National Research Nuclear University MEPhI, Moscow, 115409,  Russia}
\affiliation{
Institute of Applied Physics of the Russian Academy of Sciences, Nizhny Novgorod 603950, Russia}
\author{M.~P.~Malakhov}
\affiliation{National Research Nuclear University MEPhI, Moscow, 115409, Russia}
\affiliation{Skolkovo Institute of Science and Technology, 
Skolkovo, 121205, Russia}
\author{O.~Klimo}
\affiliation{ELI Beamlines facility, The Extreme Light Infrastructure ERIC,
Dolni Brezany 252 41, Czech Republic}
\affiliation{FNSPE, Czech Technical University in Prague, Prague, Czech Republic}
\author{S.~Weber}
\affiliation{ELI Beamlines facility, The Extreme Light Infrastructure ERIC,
Dolni Brezany 252 41, Czech Republic}

\begin{abstract} 
We reconsider the footprints of radiation friction in a head--on collision of a bunch of relativistic charged particles with a laser pulse by demonstrating that forward and backward radiation and radiation friction are coherently enhanced in a dense bunch. This opens an avenue to observe radiation friction in laser-matter interactions at much lower energies and laser intensities than accepted ever previously. An estimate for the energy losses of the particles in the bunch over the collision due to coherent radiation friction in terms of laser and bunch parameters is derived and validated by comparing with the results of three dimensional particle-in-cell simulations.  
\end{abstract}
\maketitle

{\it Introduction} -- A charged particle in an electromagnetic field accelerates, hence emits radiation, which eventually takes away its energy and momentum. In strong  fields this might have a substantial impact on the particle trajectory. Besides slowing the particle down, this effect, conventionally called radiation friction (RF) or radiation reaction \cite{dirac_prs1938,landau2,jackson_book}, can produce a highly nontrivial dynamics, including radiation trapping \cite{ji_prl2014,fedotov_pra2014,kirk_ppcf2009,gonoskov_prl2014}, the modification of the electron distribution in a bunch traversing a laser pulse \cite{tamburini_nimpra2011,chen_ppcf2010,neitz_prl2013,vranic_njp2016,niel_pre2018} and the enhancement of longitudinal plasma waves in an underdense plasma \cite{gelfer_scirep2018,gelfer_ppcf2018}. RF can be also important in practical applications, such as electron and ion laser-plasma acceleration \cite{voronin_jetp1965,zeldovich_ufn1975,dipiazza_lmp2008,tamburini_njp2010,tamburini_nimpra2011,tamburini_pre2012,chen_ppcf2010,capdessus_pre2015,gelfer_njp2021} or generation of strong magnetic fields in dense plasmas \cite{liseykina_njp2016,popruzhenko_njp2019}. It also modifies the spectrum of radiation itself \cite{koga_pop2005,dipiazza_prl2010,neitz_prl2013,thomas_prx2012}.

It has been conventionally believed that a substantial RF can arise either in a strong field or for high energy particles \cite{koga_pop2005,dipiazza_rmp2012}. Strong electromagnetic fields can be generated by high power lasers \cite{danson_hplse2019},  therefore RF attracts much attention in the context of laser--particle interactions, see, e.g., the reviews \cite{dipiazza_rmp2012,burton_contphys2014,gonoskov_rmp2022,fedotov_physrep2023}. More precisely,  
the energy of a particle is significantly altered by RF in a head--on laser--particle collision if \cite{koga_pop2005,dipiazza_lmp2008}
\begin{equation}\label{LLcond}
\mathcal{R}=\mu a_0^2\gamma\omega_L T\gtrsim1,\quad \mu=\frac{4\pi}{3}\frac{r_e}{\lambda_L},
\end{equation}
where $a_0=e E_0/m\omega_L c$ is the amplitude of the dimensionless field strength, $m$, $e$ and $\gamma$ are the particle mass, charge and Lorentz factor, $\omega_L$, $\lambda_L$ and $T$ are the pulse frequency, wavelength and duration, and $r_e=e^2/mc^2\approx 2.8\cdot 10^{-13}$~cm is the classical electron radius. For an optical laser $\mu\sim 10^{-8}$, implying that Eq.~(\ref{LLcond}) is reached with $a_0\gg1$ and/or $\gamma\gg 1$. 
 
Indeed, RF was observed experimentally \cite{cole_prx2018,poder_prx2018,los_natcomm2026} in collisions of ultrarelativistic electron beams (electron energy $\mathcal{E}\approx 0.25 - 2$~GeV,  $\gamma\approx 500-4000$) with an intense laser pulse ($a_0\approx 25$, $I\approx 1.3\cdot 10^{21}$ W/cm${}^2$, $T=45$ fs, $\lambda_L=800$ nm) ensuring $\mathcal{R}\sim 0.1-1$. Such experiments are technically challenging as require an overlap of ultrashort tightly focused laser pulses with an ultrarelativistic electron bunch and control of interaction parameters that are hard to measure directly  \cite{los_hplse2025}. Alternatively, Eq.~(\ref{LLcond}) can be met by increasing the interaction time $T$, e.g., with flying-focus pulses \cite{formanek_pra2022}.

Here we suggest another strategy to observe RF in a collision of a laser pulse with a charged particle bunch based on a coherent enhancement of bunch radiation. Indeed, while the energy of incoherent radiation of $N$ particles scales as $N$, the energy of coherent radiation scales as $N^2$ \cite{schwinger1945,michel_prl1982,hirschmugl_pra1991,hartemann_pre2000,gonoskov_pre2015,vieira_natphys2021,malaca_natphot2024,gelfer_prr2024,gelfer_mre2024,quin_ppcf2025}. Recently, we studied  coherent radiation of an electron bunch driven by a strong counterpropagating pulsed electromagnetic plane wave \cite{gelfer_prr2024}, focusing mostly on high frequencies ($\omega\gg\omega_L$). Here we estimate the main contribution to RF, show that it comes from frequencies $\omega\sim\omega_L$ and derive its dependence on the laser and particle bunch parameters. We demonstrate both analytically and numerically that coherent radiation can provide a substantial RF even at moderate particle energies and laser intensities such that inequality in Eq.~\eqref{LLcond} is violated.


In this paper we present the physical mechanism, scaling laws, and numerical validation of the coherently enhanced RF, and discuss prospects for its experimental observation. The companion paper \cite{pre_arxiv} provides a detailed derivation of the analytical model and an extended analysis of  its robustness.

{\it Coherent radiation of a particle bunch} -- Consider radiation of a bunch of charged particles colliding with a laser pulse, the latter for simplicity represented by a pulsed circularly polarized plane electromagnetic wave. Figure~\ref{fig_rad}~(a) shows the distribution of radiated electric field after the collision; see Appendix for details of the numerical approach. The radiation is directed primarily along (forward scattering) and opposite (backward scattering) the laser propagation direction. The forward scattering leaves the incident laser frequency unchanged and corresponds to a diffraction on the bunch. As discussed below, such radiation from all the particles interferes constructively, resulting in a strong coherent enhancement.

It may seem counterintuitive that near-forward radiation, emitted against the particle motion, slows particles down. Although entirely classical, the effect can be clearly explained in terms of laser photons. Before scattering, they move longitudinally; afterwards their momenta acquire transverse divergence, see Fig.~\ref{fig_rad}~(a). Since forward radiation has the incident laser frequency [see Figs.~\ref{fig_rad}~(b),(c) and \cite{pre_arxiv}], the photon energy is unchanged, so transverse momentum implies a reduced longitudinal component. The lost longitudinal momentum is partially transferred to the particles, slowing them down.


The backward radiation frequency $\omega$ is upshifted \cite{esarey_pre1993,gelfer_prr2024}, denote its wavelength $\lambda=2\pi c/\omega$. For the particles in a bunch slice of thickness $d\lesssim\lambda$ oriented perpendicular to the laser propagation direction, their backward radiation interferes constructively. If the bunch thickness $L>\lambda$, then  the emission from adjacent slices of thickness $d\sim\lambda$ interferes destructively and the total backward radiation depends on the longitudinal shape of the bunch. For example, for a Gaussian shape high frequencies with wavelengths $\lambda<L$ are suppressed exponentially \cite{hartemann_pre2000}, while for a rectangular shape (uniform density for $0<x<L$) as $\propto \omega^{-2}$ \cite{gelfer_prr2024}, see also \cite{pre_arxiv}.

To calculate radiation of a particle bunch we start from an angular--frequency distribution of the emitted radiation by $N$ particles \cite{jackson_book}
\begin{equation}\label{dE}
\begin{split}
\frac{d\mathcal{E}}{d\omega d\Omega}=
\frac{\omega^2}{c}\left|\sum\limits_{j=1}^N e_j\int\frac{dt}{2\pi}\mathbf{n}\times[\mathbf{n}\times\mathbf{v}_j]e^{i\omega\left(t-\frac{\mathbf{n}\mathbf{r}_j}{c}\right)}\right|^2,
\end{split}
\end{equation}
where $\mathbf{n}$ and $\omega$ are the direction and frequency of the emitted radiation, $e_j$, $\mathbf{r}_j(t)$ and $\mathbf{v}_j(t)$ are the charge, coordinates and velocity of the $j$-th particle of the bunch. 
It contains incoherent and coherent parts, and if all particles in the bunch have the same charges and initial velocities  (these assumptions can actually be relaxed, see below), then the coherent and incoherent parts of the radiation spectrum can be represented in the form
\cite{schiff_rsi1946,hartemann_pre2000,gelfer_prr2024,pre_arxiv}
\begin{equation}\label{cohr}
\frac{d\mathcal{E}_{coh}}{d\omega d\Omega}=\alpha N^2\frac{d\mathcal{E}^{(1)}}{d\omega d\Omega},\quad \frac{d\mathcal{E}_{incoh}}{d\omega d\Omega}=(1-\alpha) N\frac{d\mathcal{E}^{(1)}}{d\omega d\Omega} 
\end{equation}
where $d\mathcal{E}^{(1)}/d\omega d\Omega$ is the single-particle spectrum corresponding to Eq.~(\ref{dE}) with $N=1$ and $\alpha=|\left<e^{i\Phi}\right>|^2$ is the modulus squared average of the particle initial phase factor.
Coherent radiation dominates if $\alpha N\gg1$ \cite{gelfer_prr2024}; below we assume this and neglect the incoherent part. The limits $\alpha=0$ and $\alpha=1$ correspond to fully incoherent and fully coherent radiation, respectively \cite{gelfer_prr2024}.

\begin{figure*}[t]
\centering
\includegraphics[width=0.3\textwidth]{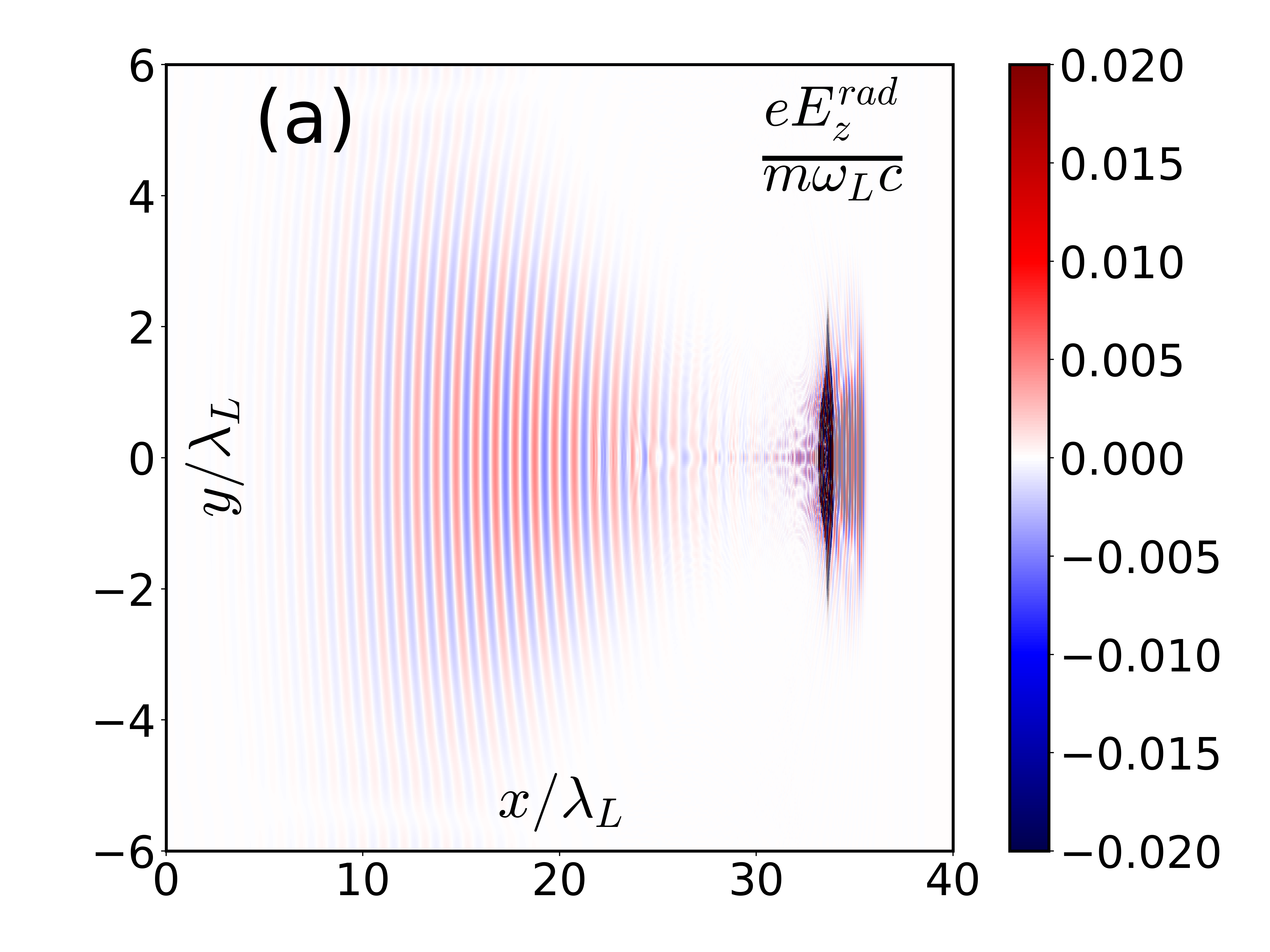}
\includegraphics[width=0.3\textwidth]{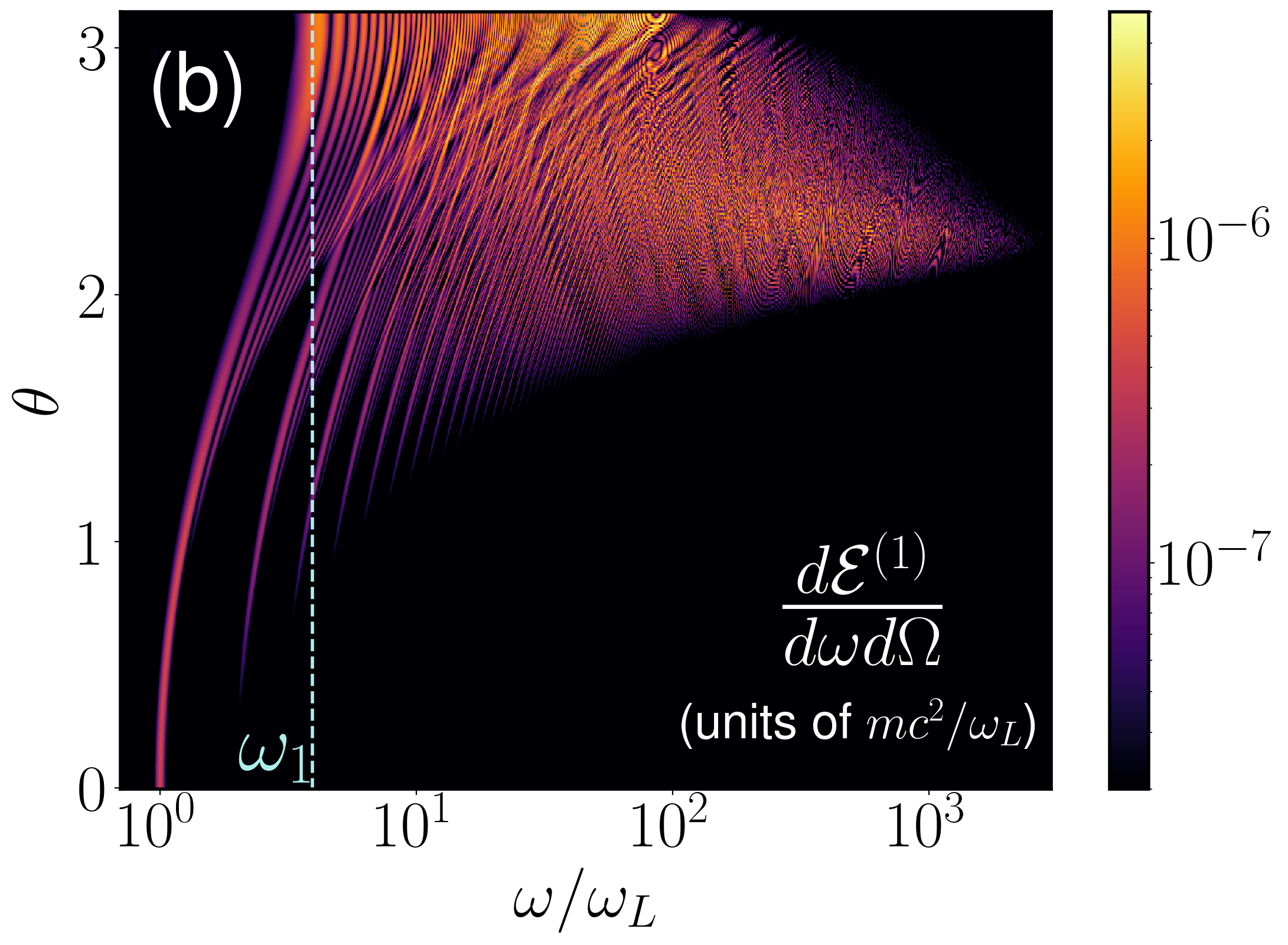}
\includegraphics[width=0.3\textwidth]{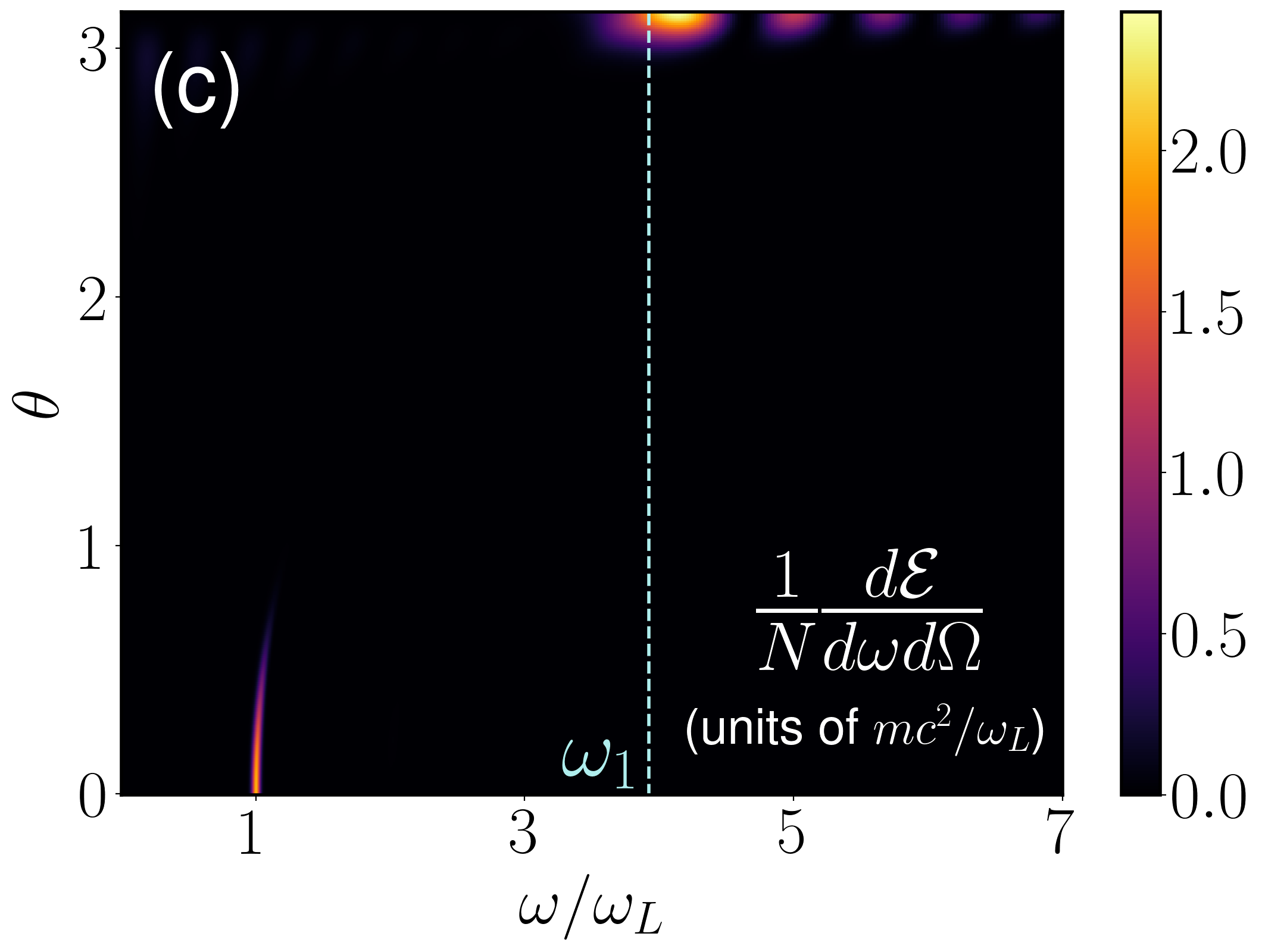}
\caption{Radiation in a head-on collision of electrons with a plane wave laser pulse. (a) -- the distributions of the $z$ component of the radiated electric field in the $(x,y)$ plane (in red and blue colors) and of the particle density (gray), the laser propagates along $x$ axis to the left; (b) -- radiation spectrum of a single particle; (c) -- per-particle spectrum of a bunch of $N\approx5\cdot10^6$ particles [see Eqs.~(\ref{cohr}), (\ref{alpha})]; $\omega_1=2\pi c/\lambda_1$, is the fundamental harmonic of the backward radiation, corresponding wavelength $\lambda_1$ is defined below Eq.~(\ref{dpbg}); $p_0=5mc$, $T\approx 50$~fs, $\lambda_L=1\mu\text{m}$.
}\label{fig_rad}
\end{figure*}

To calculate the coherence factor $\alpha$, one has to specify the spatial shape of the bunch, i.e. the density distribution $n(\mathbf{r})$ before the collision. For a Gaussian shape $n(x,\mathbf{r}_\bot)=n_0\exp\left(-\frac{x^2}{(L/2)^2}-\frac{r_\bot^2}{R^2}\right)$, where $R$ and $L$ are the transverse and longitudinal sizes of the bunch, 
\begin{equation}\label{alpha}
\alpha=e^{-\frac{\omega^2}{8c^2}\left(L^2\sin^4\frac{\theta}{2}+4R^2\sin^2\theta\right)}
\end{equation}
selects narrow coherent peaks in the forward ($\theta=0$) and, for sufficiently short bunches, backward ($\theta=\pi$) directions, for details see companion paper \cite{pre_arxiv}. Here  $\theta$ is the angle between the laser axis and the direction of the emitted radiation. Note that for a bunch  smaller than the wavelength of the emitted radiation (both $\omega R/c\ll1$ and $\omega L/c\ll1$), we have $\alpha\approx1$, meaning that all the particles in the bunch radiate coherently in all directions \cite{gonoskov_pre2015,gelfer_prr2024,quin_phd2023}. 

According to Eq.~(\ref{cohr}), radiation spectrum of the whole bunch is proportional to the single particle radiation spectrum $d\mathcal{E}^{(1)}/d\omega d\Omega$, which can be obtained by numerical evaluation 
\cite{thomas_prstab2010,boca_pra2009,seipt_pra2011}, as in  Fig.~\ref{fig_rad}~(b) \footnote{Exact analytical expressions of the single particle spectrum are known for some idealized models of the laser field, such as infinitely long monochromatic wave or in a flattop pulse of finite duration \cite{ritus1985, landau2, sarachik_prd1970, esarey_pre1993, salamin_pra1997}}. However, in forward and backward directions, where $\alpha$ peaks and therefore radiation is coherently enhanced, the analytical expression for the single particle spectrum can be evaluated approximately \cite{esarey_pre1993,seipt_lasphys2013,kharin_pra2016,gelfer_prr2024}, see companion paper \cite{pre_arxiv}. In particular, in the forward direction the particle radiates at laser frequency, while in the backward direction the radiation frequency exceeds certain threshold $\omega_1>\omega_L$. These features are clearly seen in Fig.~\ref{fig_rad}~(b) and are inherited by the radiation of the bunch, see  Figs.~\ref{fig_rad}~(a) and (c).  

{\it Momentum transfer} -- As an ultrarelativistic particle collides head-on with an ultraintense laser pulse, it initially slows down, but then reaccelerates by the ponderomotive force. In the absence of RF the net change in longitudinal momentum $p_\parallel(t)$ (the component parallel to the laser pulse propagation) over the collision is zero in accordance with the Lawson-Woodward theorem \cite{woodward_jiee1946,lawson_ieee1979}. However, RF results in a net deceleration indicated by non-vanishing $\Delta p_\parallel\equiv p_\parallel(-\infty)-p_\parallel(\infty)$, 
which can be calculated as \cite{koga_pop2005,dipiazza_lmp2008,pre_arxiv}
\begin{equation}\label{deltapll}
\Delta p_\parallel^{LL}\approx \sqrt{2\pi} \mathcal{R}p_0
\end{equation}
in accordance with Eq.~(\ref{LLcond}).  Here $p_0=p_\parallel(-\infty)$ and
the superscript $LL$ indicates that the result is derived using the Landau--Lifshitz RF force. 

Consider the average momentum loss $\left<\Delta p_\parallel\right>=\Delta P_\parallel/N$ per particle in a bunch, where $\Delta P_\parallel$ is the total momentum loss of the whole bunch. The particle momentum loss cannot be obtained by integrating the radiated energy Eq.~(\ref{cohr}) alone, since part of the emitted energy is drawn from the incident pulse. Instead we use conservation of the light-cone momentum $P_-=\mathcal{E}/c-P_\parallel$, for which the incident plane wave gives no contribution. Assuming that all the particles are ultrarelativistic during the entire course of the collision, so that $p_\parallel\approx -p_-/2$, this yields
\begin{equation}\label{dP-}
    \left<\Delta p_\parallel\right>\approx\frac{1}{2cN}\int \frac{d\mathcal{E}_{coh}}{d\omega d\Omega}(1-\cos\theta)d\omega d\Omega.
\end{equation}


Since the angular distribution peaks at $\theta=0$ and $\pi$, we split Eq.~\eqref{dP-} into forward (f, $0<\theta<\pi/2$) and backward (b, $\pi/2<\theta<\pi$) contributions. Both can be evaluated analytically when the radiation is narrow enough to use the single-particle spectra for exactly forward and backward scattering. 


The details of evaluating the integrals are given in the companion paper \cite{pre_arxiv}. As a result, the forward scattering contribution to the momentum loss reads
\begin{equation}\label{dpfg}
\begin{split}
    &\left<\Delta p^f_\parallel\right>\approx mc\frac{\pi^2}{8\sqrt{2}}\frac{n}{n_c}\frac{a_0^2}{\gamma_0^2}\omega_L T\mathcal{G},\\    &\mathcal{G}=\frac{\sqrt{A}B\left(e^{A^2}-e^{B^2}+\sqrt{\pi}B\left[\mathrm{erfi}(B)-\mathrm{erfi}(A)\right]\right)}{2\sqrt{2}\pi e^{  B^2}\sqrt{B-A}},
\end{split}
\end{equation}
where $A=\pi L^2/\left[2\lambda_L\sqrt{2(16R^2-L^2)}\right]$ and $B=4\sqrt{2}\pi R^2/(\lambda_L\sqrt{16R^2-L^2})$, $\mathrm{erfi}(z)$ is the imaginary error function, $n$ and $n_c=m\omega_L^2/4\pi e^2$ are the density of the bunch and the plasma critical density, respectively,  $\gamma_0$ is the initial Lorentz factor of the particles before the collision and it is taken into account that $N=\pi^{3/2}nR^2L/2$. 

The dependence of RF on the bunch shape is encoded in the geometric factor $\mathcal{G}$. We discuss its properties and the corresponding features of the coherent radiation in the companion paper \cite{pre_arxiv}. In particular, for a given $L\gtrsim\lambda_L$ the forward contribution Eq.~(\ref{dpfg}) gets maximal for  
\begin{equation}\label{optimum}
    \frac{R^2}{L\lambda_L}\sim {\rm const}, 
\end{equation}
where the value of the constant on the right hand side is approximately $0.05$. The corresponding maximum of the geometric factor is $\mathcal{G}^{m}\approx0.03$.

Similarly, for the backward contribution we obtain \cite{pre_arxiv}
\begin{equation}\label{dpbg}
\left<\Delta p^b_\parallel\right>\approx mc\frac{\pi\omega_L T}{32\sqrt{2}}\frac{a_0^2}{\gamma_0^2}\frac{n}{n_c}e^{-\frac{\pi^2L^2}{2\lambda_1^2}},
\end{equation}
where $\lambda_1\approx\frac{\lambda_L(1+a_0^2)}{4\gamma_0^2}$ is the wavelength of the fundamental harmonic of the backward radiation and we assumed that $2\pi L/\lambda_1\gtrsim (\omega_L T)^{1/3}\gtrsim1$. According to Eq.~(\ref{dpbg}), the backward contribution to the momentum loss is strongly suppressed for $L\gtrsim \lambda_1$. In its derivation \cite{pre_arxiv}, we assumed that $R\gtrsim\lambda_1$. Therefore, similarly to the forward contribution Eq.~(\ref{dpfg}), Eq.~(\ref{dpbg}) might be inaccurate for a small bunch with both $R,L\ll\lambda_1$. For such small bunches an accurate result can be obtained by a direct numerical evaluation of the single particle spectrum $d\mathcal{E}^{(1)}/d\omega d\Omega$ in Eq.~(\ref{dP-}), see below. Recently, RF of such small bunches was studied in Ref.~\cite{quin_prr2025} by numerical solution of the equations of motion for all particles in the bunch with  interparticle interactions fully taken into account by means of Lienard-Wiechert potentials.

{\it Numerical results} -- In order to check theoretical predictions, we performed numerical simulations using particle-in-cell (PIC) code SMILEI \cite{derouillat_cpc2018}. The details of the numerical approach can be found in the Appendix. 

\begin{figure*}[t]
\includegraphics[width=0.24\textwidth]{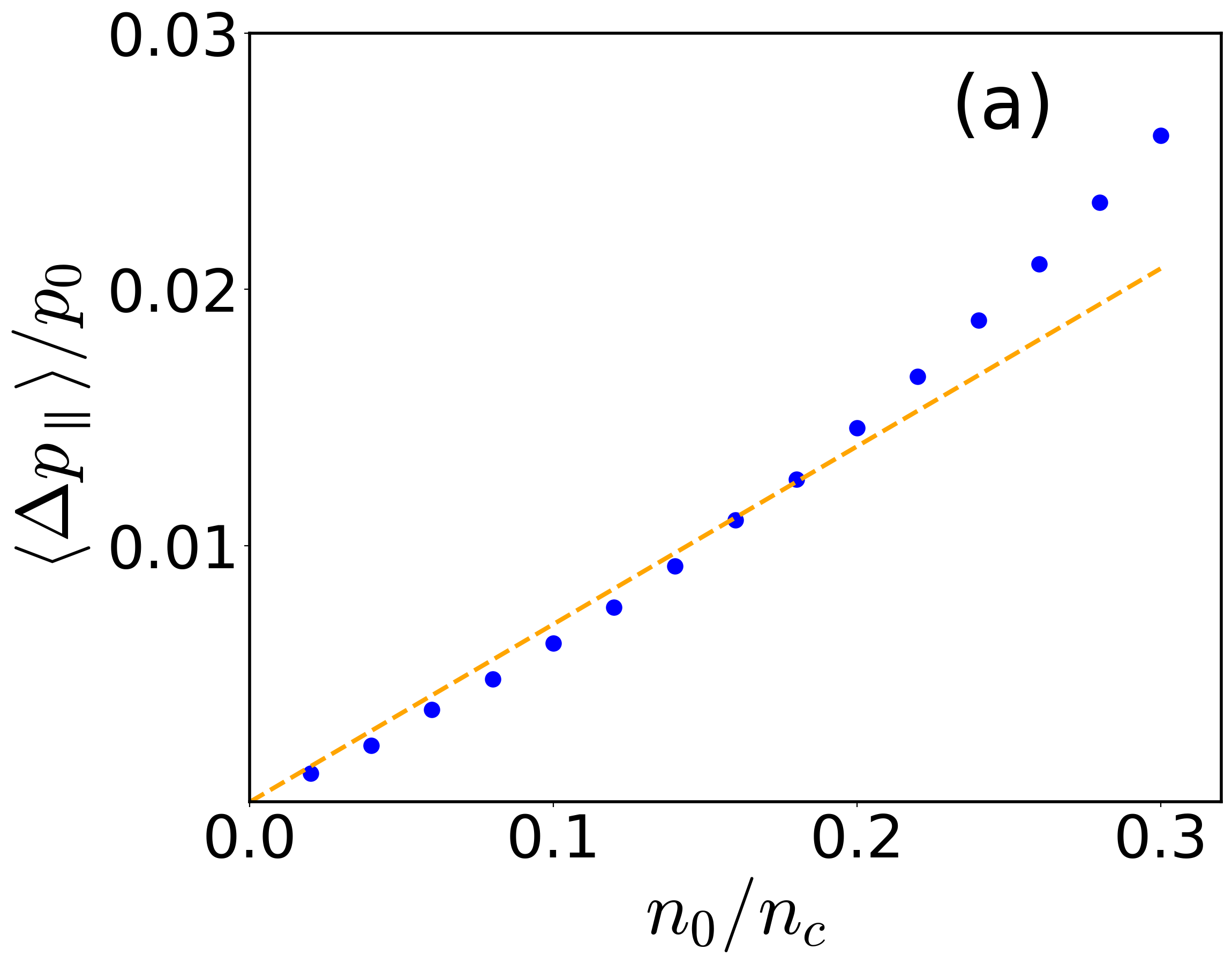}
\includegraphics[width=0.24\textwidth]{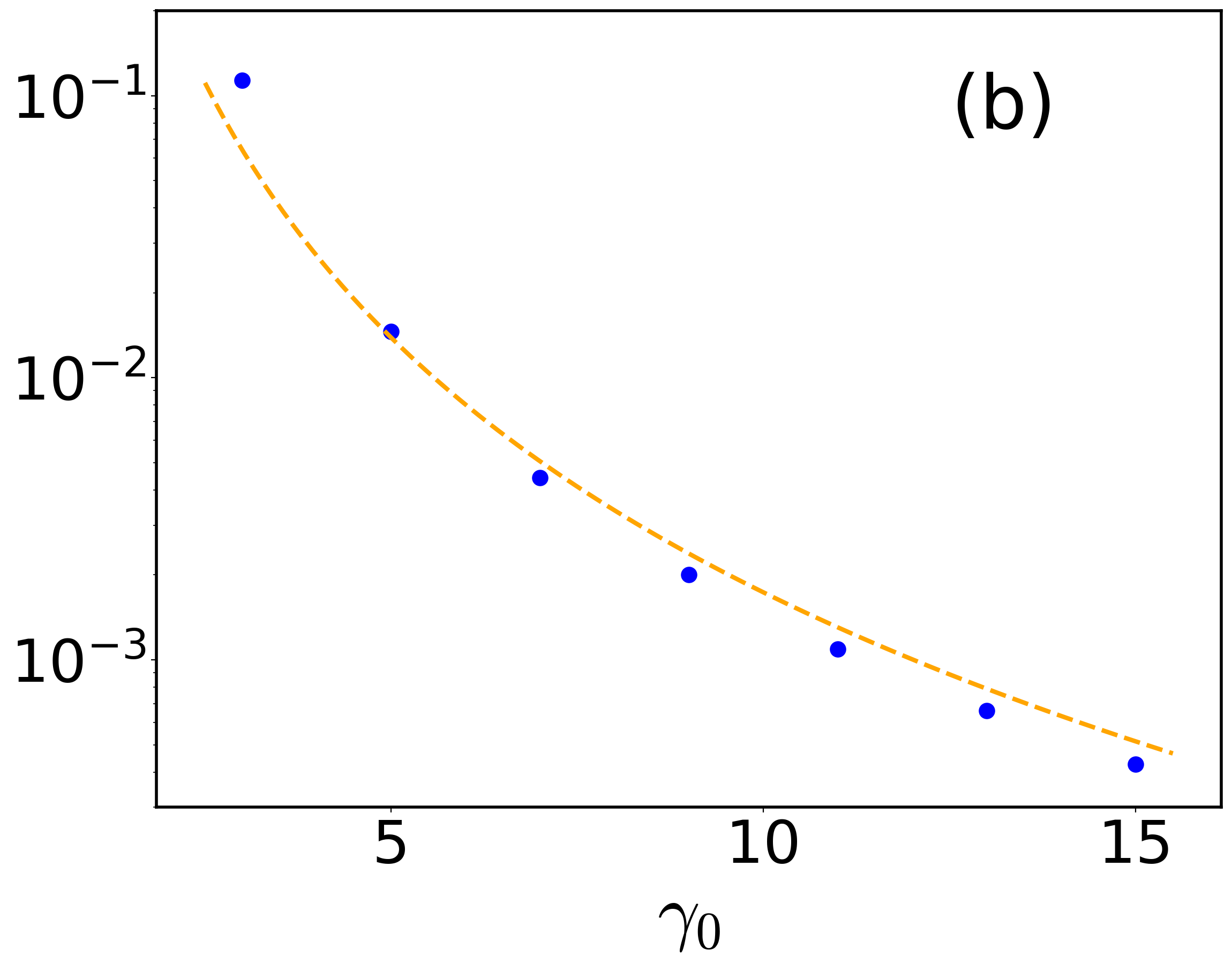}
\includegraphics[width=0.24\textwidth]{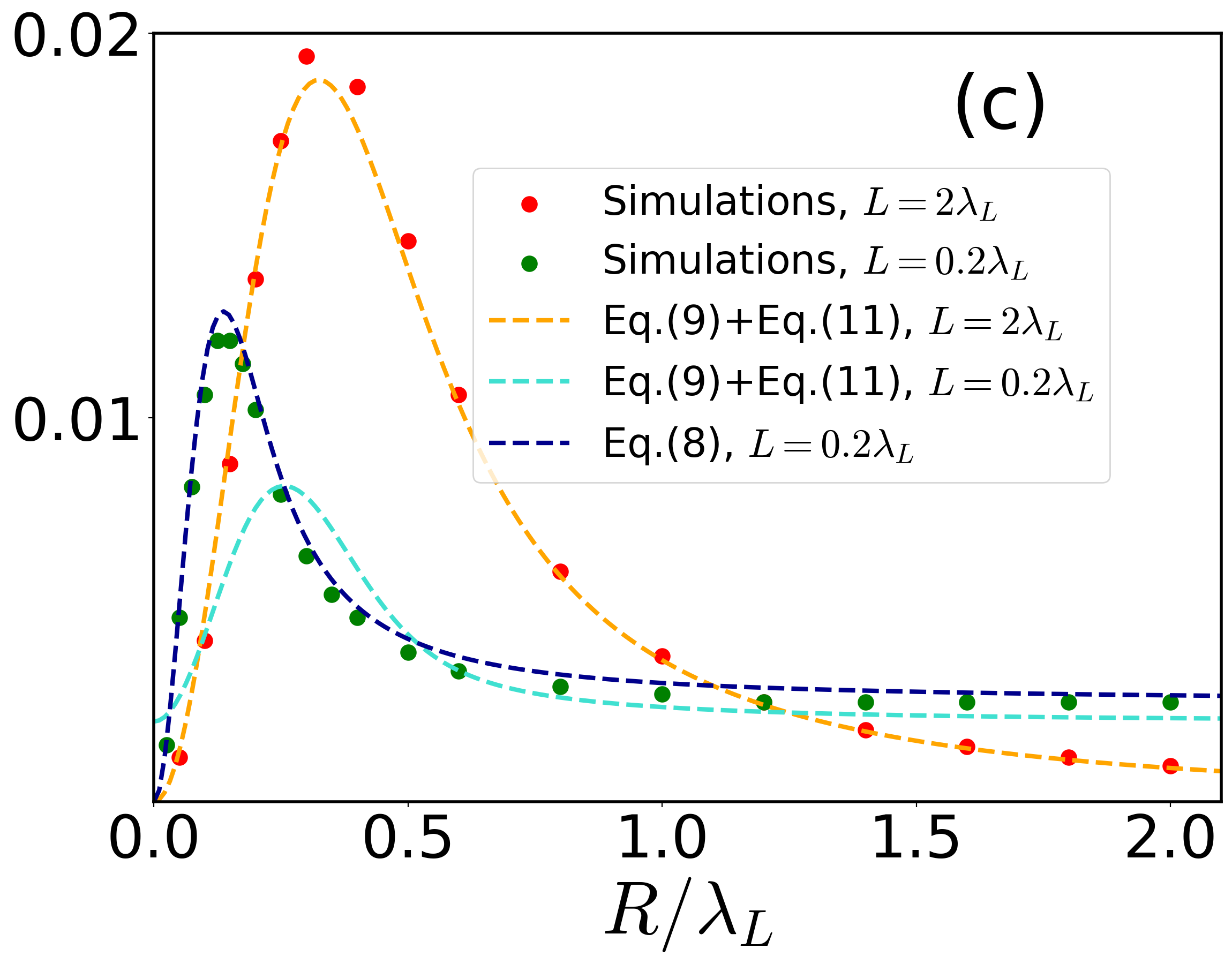}
\includegraphics[width=0.24\textwidth]{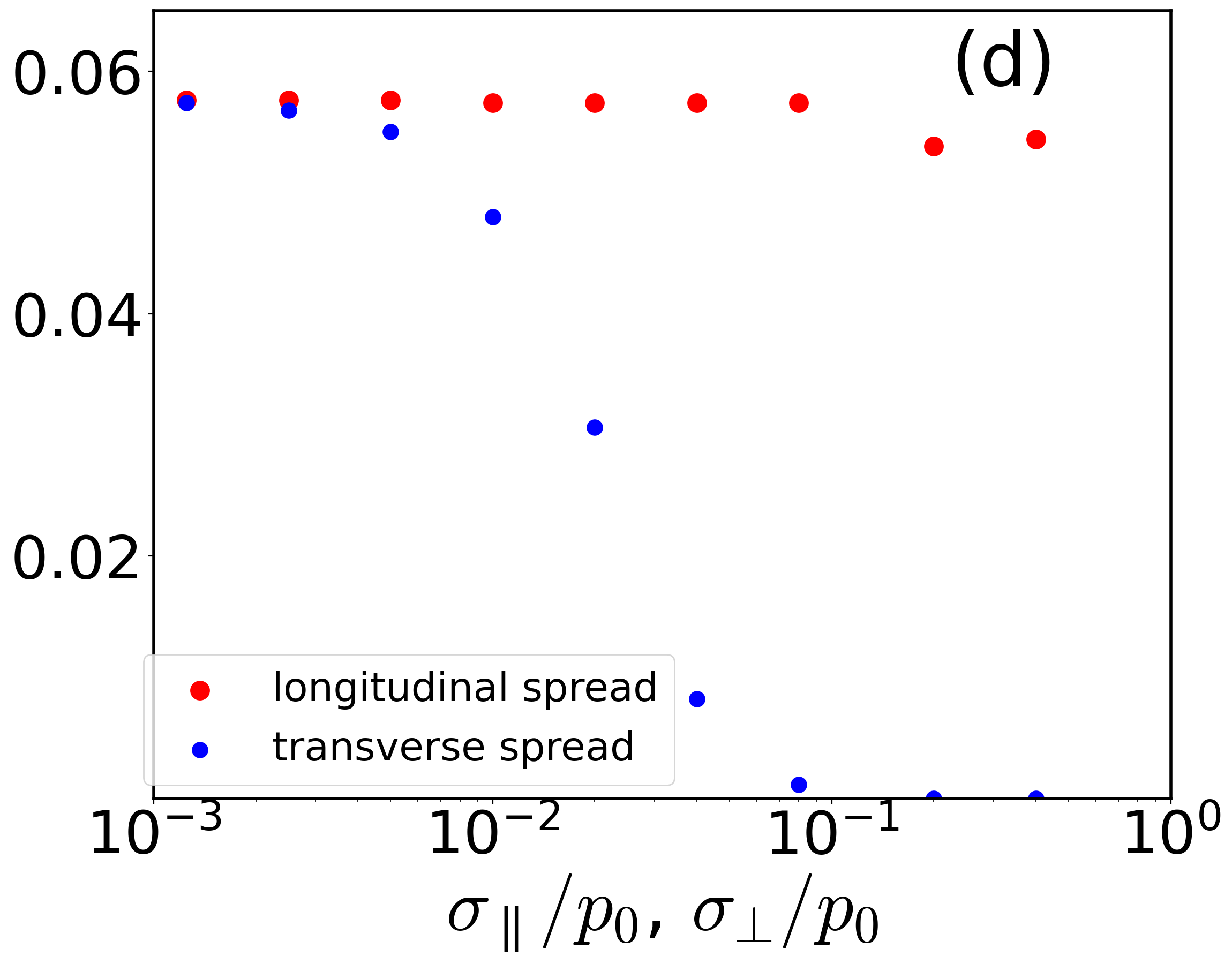}
\caption{Normalized particle deceleration $\left<\Delta p_\parallel\right>/p_0$ over the collision vs (a) bunch density $n_0$ [for $\gamma_0=5, R=0.5\lambda_L, L=2\lambda_L$], (b) initial gamma-factor $\gamma_0$ [for $n_0=0.2n_c, R=0.5\lambda_L, L=2\lambda_L$], (c) bunch width $R$ [for $n_0=0.2n_c,\gamma_0=5,L=2\lambda_L,\ 0.2\lambda_L$], and (d) bunch momentum spread [for $n=0.5 n_c$, $L=2\lambda_L$, $R=0.5\lambda_L$]. Blue, red and green dots are extracted from 3D PIC simulations, dashed curves correspond to the sum of Eqs.~(\ref{dpfg}) and (\ref{dpbg}) [orange/turquoise] and to the numerical integration of Eq.~(\ref{dP-}) with numerically calculated single particle spectrum [dark blue], red and blue dots in (d) correspond to the longitudinal spread $\sigma_\parallel$ (for $\sigma_\bot=0$) and the transverse spread $\sigma_\bot$ (for $\sigma_\parallel=0$), respectively. Other parameters: $a_0=5$, $\omega_L T=5\pi$. }\label{fig_results}
\end{figure*}

We studied the dependence of the average longitudinal momentum loss on the bunch density $n_0$, initial Lorentz factor $\gamma_0$ and the bunch sizes $R$ and $L$. 
Figure~\ref{fig_results}~(a)--(c) compares Eqs.~(\ref{dpfg}), (\ref{dpbg}) with 3D PIC simulations, showing remarkable agreement.
In Figs.~\ref{fig_results}~(a) and (b) the simulation results start to deviate from the model predictions when $\left<\Delta p_\parallel\right>/p_0\gtrsim 0.01$. This is expected as the model relies on a single particle radiation spectrum computed along the particle trajectory in a sole laser field. When RF is large, it substantially modifies the trajectories and the coherent radiation yield should be corrected accordingly. 

Figure~\ref{fig_results}~(c) illustrates the dependence of the coherent RF on the bunch width $R$ for two different values of thickness, $L=2\lambda_L$ (thick) and $L=0.2\lambda_L$ (thin). In the first case the backward contribution is negligible and Eq.~(\ref{dpfg}) [orange curve] accurately describes the effect for any $R$ including the peak value $R\sim\sqrt{0.05 L\lambda_L}\sim 0.3\lambda_L$. In the second case $L\ll\lambda_L$ the analytical estimate [turquoise curve] deviates for $R\lesssim\lambda_L$. This is natural since the assumption that radiation is sharply peaked around forward and backward directions is violated. However, by integrating Eq.~(\ref{dP-}) numerically with numerically evaluated single particle spectrum $d\mathcal{E}^{(1)}/d\omega d\Omega$, we found a very good agreement with PIC simulations even in this case, see the dark blue curve in Fig.~\ref{fig_results}~(c). At the same time, for  $R\gtrsim\lambda_L$ the sum of Eqs.~(\ref{dpfg}) and (\ref{dpbg}) is pretty accurate itself. 

The coherence of the emitted radiation might be limited by the momentum spread of the particles  \cite{angiogi_prl2018,quin_ppcf2025}. To check the robustness of our results, we performed PIC simulations with a Gaussian momentum distribution $dN/d\mathbf{p}\sim e^{-\left(p_\parallel+p_0\right)^2/2\sigma_\parallel^2-p_\bot^2/2\sigma_\bot^2}$, where $\sigma_\parallel$ and $\sigma_\bot$ are the initial longitudinal and transverse momentum spreads. Figure~\ref{fig_results}~(d) shows that RF is almost independent of the longitudinal momentum spread $\sigma_\parallel$, while the spread of the transverse momenta notably affects RF already at $\sigma_\bot\gtrsim 10^{-2} p_0$. The latter corresponds to the particle bunch divergence below $10$~mrad, which has been already achieved in acceleration of high density bunches \cite{chang_prappl2023,salehi_prx2021,storey_prstab2024,sakai_scirep2024}, see also \footnote{Note that the simulation parameters of Fig.~\ref{fig_results}~(d) [$R=0.5\lambda_L$] with the  angular spread $\delta\theta\sim10$~mrad correspond to the normalized emittance $\epsilon\sim R\delta\theta\sim10^{-2}$~mm~mrad, which is very small. However, increasing the size of the bunch keeping the parameter Eq.~\eqref{optimum} constant does not change the result for a thick bunch. Therefore the emittance is constrained mostly by the condition $R\ll w_0$, ensuring the validity of the plane wave approximation, where $w_0$ is the laser waist.}. 

A weak dependence on $\sigma_\parallel$ is natural, as for a thick bunch with $L\gtrsim\lambda_1$ the forward contribution to RF is dominant and at $\theta=0$ the particle-dependent term in the phase in Eq.~\eqref{dE} vanishes \cite{pre_arxiv}. Accordingly, the effect of momentum spread comes only from the spread in transverse components of the particle velocities in the preexponential factor of the integrand in Eq.~\eqref{dE}.

{\it Discussion} -- When a particle bunch collides with a laser pulse, the particles radiate coherently at low frequencies and incoherently at high frequencies \cite{gelfer_prr2024}. For $a_0\gg1$ the incoherent radiation comes at $\omega\simeq a_0 (a_0^2+\gamma^2)\omega_L\gg \omega_L$ similarly to the single particle case \cite{jackson_book,ritus1985,esarey_pre1993}, see  Fig.~\ref{fig_rad}~(b). In contrast, the coherent radiation is peaked at $\omega\sim\omega_L$ in the forward and at $\omega\sim\omega_1$ in the backward direction, respectively, and can be orders of magnitude stronger than the incoherent one, cf. Figs.~\ref{fig_rad}~(c) and (b).

Hence, for $a_0\gg1$ the coherent and incoherent contributions to RF are separated in the frequency domain. They can be compared using Eqs.~(\ref{deltapll}) and (\ref{dpfg}). For the optimal bunch shape [see Eq.~(\ref{optimum})] we have
\begin{equation}\label{ceqll}
\frac{\left<\Delta p_\parallel\right>}{\Delta p_\parallel^{LL}}\sim \frac{n}{n_c}\frac{\mathcal{G}^m}{\mu\gamma_0^4}, 
\end{equation}
where $\mu$ is introduced in Eq.~(\ref{LLcond}) and $\mathcal{G}^m\approx0.03$. Since for an optical laser $\mu\sim 10^{-8}$, for a mildly relativistic bunch the coherent contribution should dominate already at bunch density well below the critical one. 

\begin{figure}[t]
\includegraphics[width=0.48\textwidth]{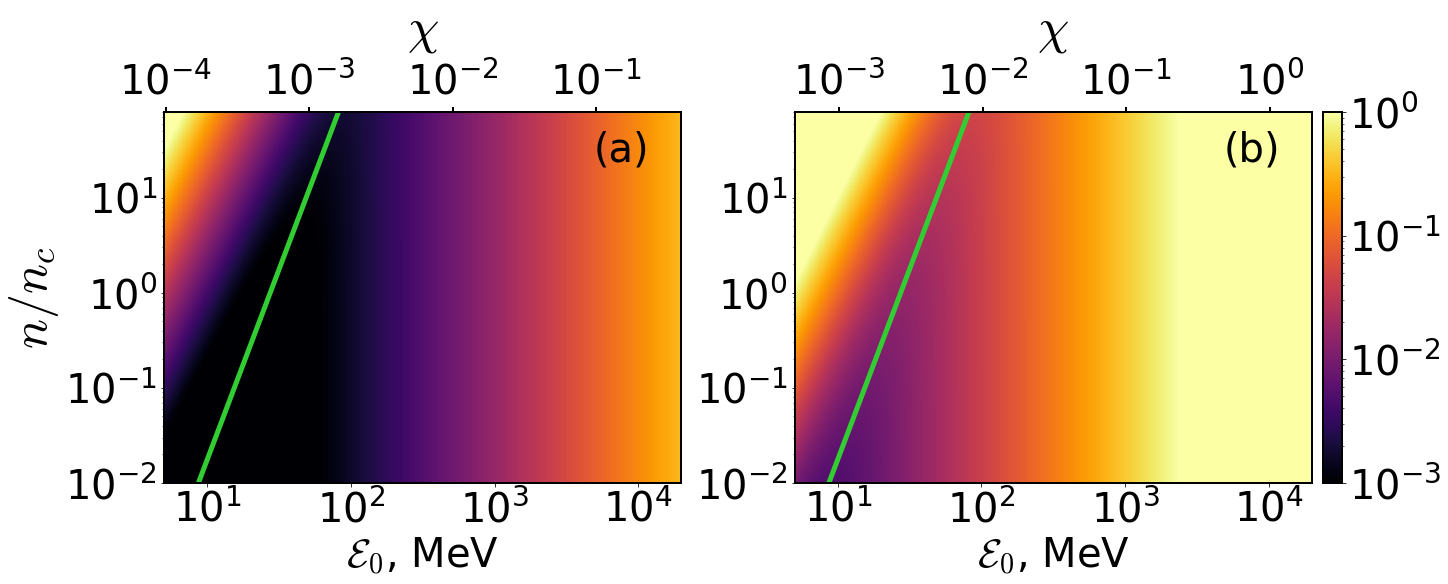}
\caption{Normalized particle deceleration $\frac{\left<\Delta p_\parallel\right>+\Delta p_\parallel^{LL}}{p_0}$ due to the coherent $\left<\Delta p_\parallel\right>$  and incoherent $\Delta p^{LL}_\parallel$ contributions to RF for (a) $a_0=5$ [$I\approx 6.6\cdot 10^{19}$ W/cm$^2$] and  (b) $a_0=25$ [$I\approx 1.7\cdot 10^{21}$ W/cm$^2$]. The green line corresponds to $\left<\Delta p_\parallel\right>\sim\Delta p^{LL}_\parallel$, see Eq.~(\ref{ceqll}). Other parameters: $\omega_L T=15\pi$ [FWHM$\approx50$fs], $\lambda_L=1\mu$m.}\label{fig_rfest}
\end{figure}

The dependence of the magnitude of RF on the bunch density and initial energy is illustrated in Fig.~\ref{fig_rfest}, where the ratio of the sum $\left<\Delta p_\parallel\right>+\Delta p_\parallel^{LL}$ to the initial momentum $p_0$ is shown for the optimal bunch shape. At a lower laser intensity [$a_0=5, I\approx 6.6\cdot 10^{19}$ W/cm$^2$, see Fig.~\ref{fig_rfest}~(a)] RF is substantial and potentially observable in two well separated regions at the top left and right corners of the plot. For lower bunch energy the major contribution to RF comes from coherent radiation at low frequencies and is proportional to the bunch density, while for a higher bunch energy it comes from incoherent radiation at high frequencies and is density independent.  

At higher intensity  [$a_0=25, I\approx 1.7\cdot 10^{21}$ W/cm$^2$, see Fig.~\ref{fig_rfest}~(b)] RF remains substantial when the coherent and incoherent contributions are comparable (tilted green line). Here, however, one cannot expect Eqs.~(\ref{deltapll}) and (\ref{dpfg}) to be accurate and that the total RF might be combined as their simple sum, as their derivations assume that only one contribution dominates. Their joint regime is a separate interesting problem that we address for future studies. 

Let us discuss the prospects for experimental observation of the coherently enhanced RF. The required intensity $I\gtrsim 10^{18}-10^{20}$ W/cm$^2$ corresponding to $a_0\sim1-10$ and the required particle energy $10-100$ MeV are now routinely obtained at the modern laser facilities \cite{danson_hplse2019}. However, production of  relativistic particle bunches of the required high densities might be really challenging. Obtaining electron bunches with densities of the order of $10^{18}-10^{19}$ cm$^{-3}$, low emittance and energy spread $\delta\mathcal{E}/\mathcal{E}\sim1-20\%$ using laser/plasma wakefield accelerators were reported recently \cite{salehi_prx2021,chang_prappl2023,storey_prstab2024}. Maintaining a small energy spread is essential, because $\left<\Delta p_\parallel\right>/p_0\gtrsim\delta\mathcal{E}/\mathcal{E}$ is necessary for a straightforward observation of the effect. An even higher density $n\sim 10^{21}$ cm$^{-3}$ is anticipated at the the FACET~II facility in the near future \cite{yakimenko_prl2019,facet2} and in schemes based on electron acceleration by a helical laser \cite{shi_prl2021,shi_ppcf2021,shi_hplse2022,blackman_commphys2022}. For an optical laser ($\lambda_L\sim1\mu$m) these densities correspond to $\left(10^{-3}-1\right)n_c$. For electron-positron bunches the state-of-the-art densities $10^{16}$ cm$^{-3}$  \cite{sarri_ncomms2015} correspond to $10^{-5} n_c$ for an optical laser.

Note, however, that the ratio $n/n_c$ could be also increased by lowering $n_c$, i.e. by using a larger wavelength infrared laser instead of an optical one. For example, the critical plasma density  corresponding to a $10 \mu$m $\mathrm{CO}_2$ laser is 100 times less than that for a $1\mu$m optical laser. The power of $\mathrm{CO}_2$ lasers has already reached $10$ TW level \cite{haberberger_optexpr2010,polyanskiy_osacont2020,chang_advoptphot2022}, thus making accessible the required here mildly relativistic regime $1<a_0<10$. 

It was initially unclear if radiation
in experiments \cite{cole_prx2018,poder_prx2018} was emitted by electrons in a classical or quantum regime. Only recently by collecting additional data the quantum regime of RF for that setup was confirmed~\cite{los_natcomm2026}. In fact, observation of purely classical regime of RF in such experiments would be challenging, since the  parameter Eq.~(\ref{LLcond}) ruling the significance of RF is proportional to the quantum  parameter $\chi\approx2\gamma a_0\hbar\omega_L/mc^2$ [$\hbar$ is the Planck constant], which governs the importance of quantum effects \cite{ritus1985,dipiazza_rmp2012}. For typical parameters of modern high power laser facilities \cite{danson_hplse2019} $a_0\sim 10-100$ and $\omega_L T\sim 100$, the condition $\mathcal{R}\gtrsim 1$  implies $\chi\gtrsim 10^{-2}$ [see also Fig.~\ref{fig_rfest}], indicating that quantum effects are non-negligible. Observation of RF with mildly relativistic dense coherently radiating bunches instead of ultrarelativistic rarified ones requires much smaller values $\chi\sim10^{-4}-10^{-3}$, thus giving access to a yet experimentally unexplored purely classical regime of RF.

To conclude, dense particle bunches can lose substantial energy and momentum in laser-bunch collisions via coherent radiation. Here we described analytically the dependence of the momentum loss by such a dense particle bunch colliding head--on with a laser pulse on the relevant laser and bunch parameters. This is a first quantitative prediction for the scaling laws of coherent RF. Three dimensional PIC simulations confirm a high accuracy of our model. We emphasize that the coherently enhanced RF under consideration is not a correction to an ``ordinary'' RF, which has been studied theoretically for decades and observed on the experiments \cite{poder_prx2018,cole_prx2018,los_natcomm2026}, but  rather is a new mechanism of RF operating in a different range of parameters. The major contribution to coherent RF comes from forward and backward scattered radiation at low frequencies. Our findings demonstrate that  observation of RF with dense particle bunches might be possible even at moderate particle energies and laser intensities. The coherent regime of radiation might be relevant also to the ongoing discussions of backreaction of the emitted radiation on the driving laser pulse \cite{fedotov_prl2010,seipt_prl2017}, but we leave this for future studies.


{\it Acknowledgments} -- The authors are grateful to Antonino Di Piazza, Matteo Tamburini, Mickael Grech, Caterina Riconda, Sergey Rykovanov, Igor Kostyukov, Vladimir Tikhonchuk, Martin Stack Formanek and Peter Valenta for valuable discussions. E.G.G., O.K. and S.W. were supported by NSF--GACR project 24-14395L. A.M.F. was supported by the Russian Science Foundation (Grant No. 25-12-00336). The numerical simulations were performed using the code SMILEI and the resources of the ELI ERIC SUNRISE cluster. 
\appendix
\section*{Appendix. Numerical approach}


In a PIC simulation, at each time step the coordinates and momenta of macroparticles are updated according to the equations of motion in an instant electromagnetic field. In turn, charge and current densities of macroparticles after smearing out across the cells serve as instant sources in the Maxwell equations governing the field evolution over the time step \cite{dawson_rmp1983,birdsall_book2018}. 

The account for RF is conventionally taken by means of two major approaches \cite{gonoskov_pre2015}. The first one is just to add a RF force term to the RHS of the equation of motion for a macroparticle. This is supposed reasonable when the quantum  parameter
\begin{equation}\label{chi0}
\chi\sim\frac{\gamma E}{E_{cr}},
\end{equation}
which measures the ratio of the electric field $E$ in the particle rest frame to the critical field of quantum electrodynamics (QED) $E_{cr}=m^2 c^3/e\hbar$  \cite{sauter1931,schwinger1951}, is so small that the quantum effects of radiation can be neglected \cite{ritus1985,dipiazza_rmp2012}.

To account for quantum effects (including, e.g., struggling \cite{shen_prl1972,blackburn_prl2014} and quenching \cite{harvey_prl2017}), instead of adding a classical RF force, one subdivides photons into soft and hard according to their energy, the latter also represented by macroparticles \cite{elkina_prstab2011}. Their emission at each time step is then implemented as a random splitting off from charged macroparticles according to the probability distributions for photon emission in a locally constant crossed field approximation \cite{ritus1985,gonoskov_pre2015,dipiazza_pra2021,gelfer_prd2022}. 
This way RF emerges as a quantum recoil of charged macroparticles  ruled by energy-momentum conservation in a sequence of hard photon emissions. Soft photons are commonly ignored as  their contribution to the ordinary (incoherent) RF can be neglected for strong fields ($a_0\gtrsim1$) and ultrarelativistic particles ($\gamma\gg1$).

\begin{figure}[t]
\includegraphics[width=0.23\textwidth]{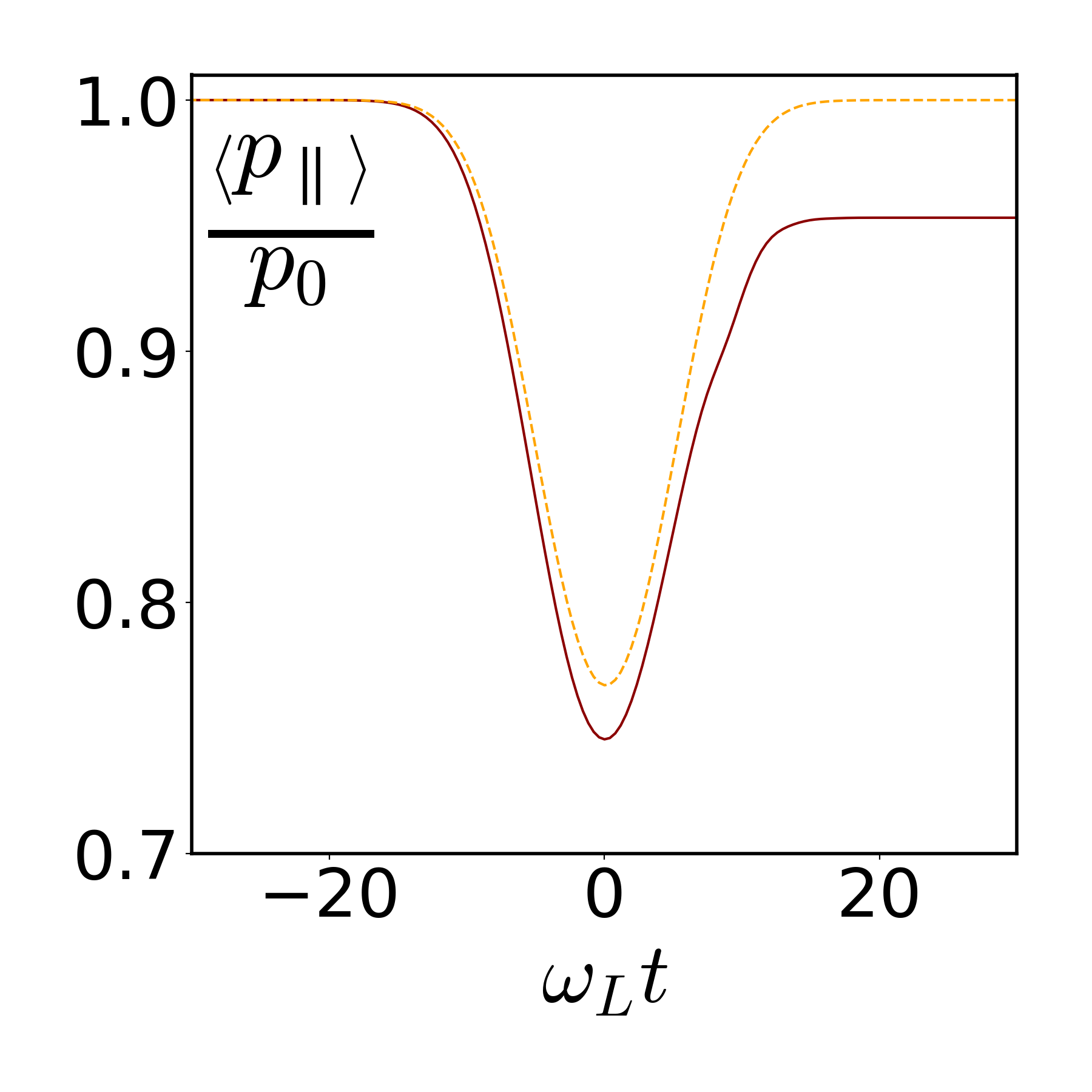}
\includegraphics[width=0.23\textwidth]{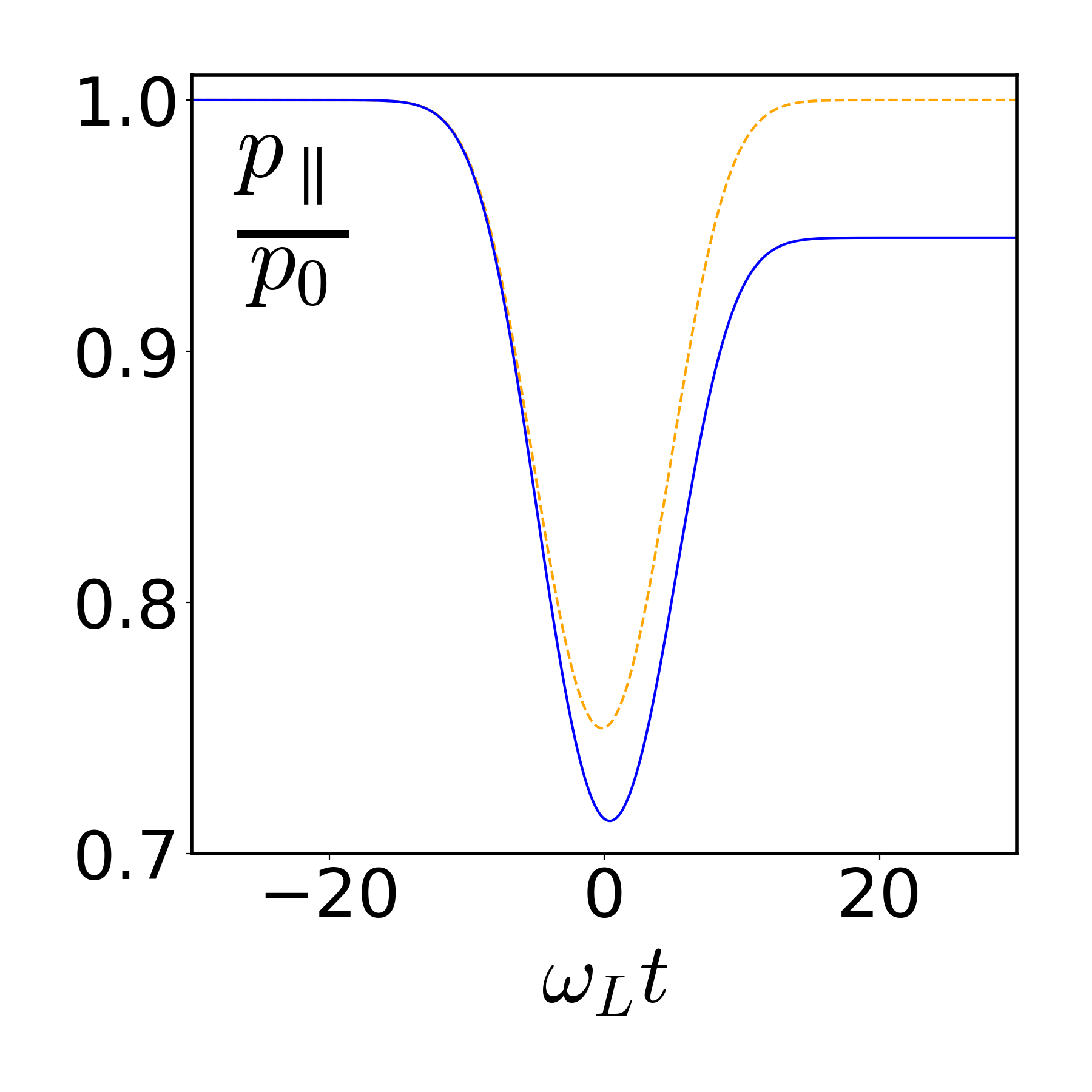}
\caption{(a) -- Evolution of the average longitudinal momentum of particles in a neutral electron--positron bunch colliding with a laser pulse obtained by numerical simulations for a dense $n=0.4n_c$, $N\approx1.7\cdot10^8$ (solid dark red line) and rarefied $n=4\cdot 10^{-5}n_c$, $N\approx 1.7\cdot 10^4$ (dashed orange line) bunch for $a_0=\gamma_0=5$, $T\approx 17$ fs, $R=0.3\lambda_L$, $L=1.5\lambda_L$, $\lambda_L=1\mu \text{m}$; (b) -- evolution of the longitudinal momentum of a charged particle colliding with a laser pulse with (solid blue line) and without (dashed orange line) account for RF for $a_0=\gamma_0=50$, $T\approx 17$ fs, $\lambda_L=1\mu\text{m}$.}\label{fig_ll}
\end{figure}

As they stand, both approaches  account only for an incoherent contribution to RF. However, a coherent contribution also partially reveals in  PIC simulations along with those part of classical radiation of charged macroparticles that is resolved on the grid, i.e. those with wavelengths $\lambda\gtrsim d$, where $d$ is the cell size  \cite{gonoskov_pre2015}. This part of radiation correctly accounts for relative phases and hence for the coherence of radiating charged macroparticles, as in Eq.~(\ref{dE}) \cite{gonoskov_pre2015,gelfer_prr2024}. With this RF emerges as a transfer of energy and momentum from macroparticles to their (coherent) radiation. This is seen in Fig.~\ref{fig_ll}~(a), where we compare simulation of the evolution of the average particle longitudinal momentum in the bunches of the same size but different densities with both the Landau-Lifshitz RF force and the quantum photon emission modules of the PIC code switched off. For dense particle bunches such contribution to RF associated with coherent radiation at low frequencies $\omega\sim\omega_L$  can be essential and even dominant. 

Note that qualitatively the evolution of the average momentum of a particle in a bunch is similar to the evolution of the momentum of a single particle subjected to Landau-Lifshitz RF force, cf. Figs.~\ref{fig_ll}~(a) and (b). However, it is worth emphasizing that due to coherent enhancement of radiation the particles in a bunch can lose a comparable amount of momentum for much lower values of energy and intensity.

Dense relativistic electron bunch creates a strong self-field. This field increases the momentum spread in the bunch, but should not affect its total momentum, hence the average particle momentum. However, unless the solution of the relativistic Poisson equation at the initialization step of a simulation is perfectly accurate, a self-field can accelerate or decelerate the bunch as a whole and this is what we indeed observed in the simulations. To overcome the issue, we had to switch to simulations of neutral electron-positron bunches, for which such a numerical artifact is suppressed. 

Fortunately, our approach works for bunches constituted of two species with the same masses and opposite charges, $|e_j|=e$. Indeed, in a circularly polarized field oppositely charged particles of equal masses orbit in the transverse plane with the same magnitude of velocity but in opposite directions. Therefore, for forward and backward scattering that make a major contribution to RF, the factor $e_j[\mathbf{n}\times[\mathbf{n}\times \mathbf{v}_j(t)]]$ in Eq.~(\ref{dE}) is actually the same for all the particles in such a bunch, thus justifying  Eq.~(\ref{cohr}) and all the following. Hence, our theoretical predictions are equally valid for neutral electron-positron bunches, for which we compare them to the results of 3D numerical simulations. 

In the simulations all the electrons and positrons constituting the bunch are moving from left to right with the same initial momenta [except the case of Fig.~\ref{fig_results}~(d)], while the plane wave circularly polarized laser pulse is moving from right towards the bunch. The RF modules of the PIC code are switched off because the contribution of incoherent RF is negligible for the chosen simulation parameters and the deceleration is provided solely by the coherent radiation resolved on the grid. We extract from simulations a variation of the particle average momentum over the collision. The initial densities of electrons and positrons are given by the expression above Eq.~(\ref{alpha}) with the only replacement $n_0\to n_0/2$ (so that $n_0$  now stands for the maximal value of the total particle density in the bunch). Each cell with plasma contains 5 macroparticles of each kind and the simulation box size is $25\lambda_L\times 8\lambda_L\times 8\lambda_L$. The spatial resolution is $64$ cells per laser wavelength and the temporal resolution is $128$ time steps per laser period. This is enough to resolve the frequencies of the coherent radiation contributing to RF. The transverse and longitudinal boundary conditions are periodic and absorptive, respectively. 

\bibliography{lit_arxiv}

@article{danson_hplse2019,
    author = {Danson, Colin N and Haefner, Constantin and Bromage, Jake and Butcher, Thomas and Chanteloup, Jean-Christophe F and Chowdhury, Enam A and Galvanauskas, Almantas and Gizzi, Leonida A and Hein, Joachim and Hillier, David I and others},
    doi = {10.1017/hpl.2019.36},
    journal = {High Power Laser Sci. Eng.},
    publisher = {Cambridge University Press},
    title = {Petawatt and exawatt class lasers worldwide},
    volume = {7},
    year = {2019}
}

@article{dipiazza_rmp2012,
    author = {Di Piazza, A and M{\"u}ller, C and Hatsagortsyan, KZ and Keitel, Ch H},
    doi = {10.1103/RevModPhys.84.1177},
    journal = {Rev. Mod. Phys.},
    number = {3},
    pages = {1177},
    publisher = {APS},
    title = {Extremely high-intensity laser interactions with fundamental quantum systems},
    volume = {84},
    year = {2012}
}

@article{elkina_prstab2011,
    author = {Elkina, N V and Fedotov, A M and Kostyukov, I Yu and Legkov, M V and Narozhny, N B and Nerush, E N and Ruhl, H},
    doi = {10.1103/PhysRevSTAB.14.054401},
    journal = {Phys. Rev. STAB},
    number = {5},
    pages = {054401},
    publisher = {APS},
    title = {{QED} cascades induced by circularly polarized laser fields},
    volume = {14},
    year = {2011}
}

@article{fedotov_prl2010,
    author = {Fedotov, A M and Narozhny, N B and Mourou, G{\'e}rard and Korn, Georg},
    doi = {10.1103/PhysRevLett.105.080402},
    journal = {Phys. Rev. Lett.},
    number = {8},
    pages = {080402},
    publisher = {APS},
    title = {Limitations on the attainable intensity of high power lasers},
    volume = {105},
    year = {2010}
}

@article{fedotov_pra2014,
    author = {Fedotov, A M and Elkina, N V and Gelfer, E G and Narozhny, N B and Ruhl, H},
    doi = {10.1103/PhysRevA.90.053847},
    journal = {Phys. Rev. A},
    number = {5},
    pages = {053847},
    publisher = {APS},
    title = {Radiation friction versus ponderomotive effect},
    volume = {90},
    year = {2014}
}

@book{landau2,
    author = {Landau, Lev Davidovich and Lifshitz, I M},
    publisher = {Course of Theoretical Physics Series, Pergamon Press, London},
    title = {Theoretical Physics: The Classical Theory of Fields},
    year={1988},
    volume = {2}
}

@article{ritus1985,
  	title={Quantum effects of the interaction of elementary particles with an intense electromagnetic field},
	author={Ritus, V I},
	journal={J. Russ. Laser Res.},
	volume={6},
	number={5},
	pages={497--617},
	year={1985},
	publisher={Springer},
	doi={https://doi.org/10.1007/BF01120220}
	}

@article{schwinger1951,
	title={On gauge invariance and vacuum polarization},
	author={Schwinger, Julian},
	journal={Phys. Rev.},
	volume={82},
	number={5},
	pages={664},
	year={1951},
	publisher={APS},
	doi={10.1103/PhysRev.82.664}
}

@article{dipiazza_pra2021,
  title = {Unveiling the transverse formation length of nonlinear {C}ompton scattering},
  author = {Di Piazza, A.},
  journal = {Phys. Rev. A},
  volume = {103},
  issue = {1},
  pages = {012215},
  numpages = {10},
  year = {2021},
  month = {Jan},
  publisher = {American Physical Society},
  doi = {10.1103/PhysRevA.103.012215},
  url = {https://link.aps.org/doi/10.1103/PhysRevA.103.012215}
}

@article{niel_pre2018,
  title = {From quantum to classical modeling of radiation reaction: A focus on stochasticity effects},
  author = {Niel, F. and Riconda, C. and Amiranoff, F. and Duclous, R. and Grech, M.},
  journal = {Phys. Rev. E},
  volume = {97},
  issue = {4},
  pages = {043209},
  numpages = {27},
  year = {2018},
  month = {Apr},
  publisher = {American Physical Society},
  doi = {10.1103/PhysRevE.97.043209},
  url = {https://link.aps.org/doi/10.1103/PhysRevE.97.043209}
}

@article{cole_prx2018,
  title = {Experimental Evidence of Radiation Reaction in the Collision of a High-Intensity Laser Pulse with a Laser-Wakefield Accelerated Electron Beam},
  author = {Cole, J. M. and Behm, K. T. and Gerstmayr, E. and Blackburn, T. G. and Wood, J. C. and Baird, C. D. and Duff, M. J. and Harvey, C. and Ilderton, A. and Joglekar, A. S. and others},
  journal = {Phys. Rev. X},
  volume = {8},
  issue = {1},
  pages = {011020},
  numpages = {11},
  year = {2018},
  month = {Feb},
  publisher = {American Physical Society},
  doi = {10.1103/PhysRevX.8.011020},
  url = {https://link.aps.org/doi/10.1103/PhysRevX.8.011020}
}

@article{poder_prx2018,
  title = {Experimental Signatures of the Quantum Nature of Radiation Reaction in the Field of an Ultraintense Laser},
 author = {Poder, K. and Tamburini, M. and Sarri, G. and Di Piazza, A. and Kuschel, S. and Baird, C. D. and Behm, K. and Bohlen, S. and Cole, J. M. and Corvan, D. J. and others},
  journal = {Phys. Rev. X},
  volume = {8},
  issue = {3},
  pages = {031004},
  numpages = {11},
  year = {2018},
  month = {Jul},
  publisher = {American Physical Society},
  doi = {10.1103/PhysRevX.8.031004},
  url = {https://link.aps.org/doi/10.1103/PhysRevX.8.031004}
}

@article{derouillat_cpc2018,
  title={Smilei: A collaborative, open-source, multi-purpose particle-in-cell code for plasma simulation},
  author={Derouillat, Julien and Beck, Arnaud and P{\'e}rez, F and Vinci, Tommaso and Chiaramello, M and Grassi, Anna and Fl{\'e}, M and Bouchard, G and Plotnikov, I and Aunai, Nicolas and others},
  journal={Comput. Phys. Comm.},
  volume={222},
  pages={351--373},
  year={2018},
  doi={10.1016/j.cpc.2017.09.024},
  publisher={Elsevier}
}

@article{sauter1931,
  title={{\"U}ber das {V}erhalten eines {E}lektrons im homogenen elektrischen {F}eld nach der relativistischen {T}heorie {D}iracs},
  author={Sauter, Fritz},
  journal={Z. Phys.},
  volume={69},
  number={11},
  pages={742--764},
  year={1931},
  doi={10.1007/BF01339461},
  publisher={Springer}
}

@article{gonoskov_pre2015,
  title = {Extended particle-in-cell schemes for physics in ultrastrong laser fields: Review and developments},
  author = {Gonoskov, A. and Bastrakov, S. and Efimenko, E. and Ilderton, A. and Marklund, M. and Meyerov, I. and Muraviev, A. and Sergeev, A. and Surmin, I. and Wallin, E.},
  journal = {Phys. Rev. E},
  volume = {92},
  issue = {2},
  pages = {023305},
  numpages = {18},
  year = {2015},
  month = {Aug},
  publisher = {American Physical Society},
  doi = {10.1103/PhysRevE.92.023305},
  url = {https://link.aps.org/doi/10.1103/PhysRevE.92.023305}
}

@article{fedotov_physrep2023,
  title={Advances in QED with intense background fields},
  author={Fedotov, A and Ilderton, A and Karbstein, F and King, Ben and Seipt, D and Taya, H and Torgrimsson, Greger},
  journal={Phys. Rep.},
  volume={1010},
  pages={1--138},
  year={2023},
  publisher={Elsevier}
}

@article{sarachik_prd1970,
  title = {Classical Theory of the Scattering of Intense Laser Radiation by Free Electrons},
  author = {Sarachik, E. S. and Schappert, G. T.},
  journal = {Phys. Rev. D},
  volume = {1},
  issue = {10},
  pages = {2738--2753},
  numpages = {0},
  year = {1970},
  month = {May},
  publisher = {American Physical Society},
  doi = {10.1103/PhysRevD.1.2738},
  url = {https://link.aps.org/doi/10.1103/PhysRevD.1.2738}
}

@article{salamin_pra1997,
  title = {Harmonic generation by scattering circularly polarized light of arbitrary intensity from free electrons of arbitrary initial velocity},
  author = {Salamin, Yousef I. and Faisal, Farhad H. M.},
  journal = {Phys. Rev. A},
  volume = {55},
  issue = {5},
  pages = {3964--3967},
  numpages = {0},
  year = {1997},
  month = {May},
  publisher = {American Physical Society},
  doi = {10.1103/PhysRevA.55.3964},
  url = {https://link.aps.org/doi/10.1103/PhysRevA.55.3964}
}

@article{angiogi_prl2018,
  title = {Quantum Limitation to the Coherent Emission of Accelerated Charges},
  author = {Angioi, A. and Di Piazza, A.},
  journal = {Phys. Rev. Lett.},
  volume = {121},
  issue = {1},
  pages = {010402},
  numpages = {6},
  year = {2018},
  month = {Jul},
  publisher = {American Physical Society},
  doi = {10.1103/PhysRevLett.121.010402},
  url = {https://link.aps.org/doi/10.1103/PhysRevLett.121.010402}
}

@article{gonoskov_rmp2022,
  title = {Charged particle motion and radiation in strong electromagnetic fields},
  author = {Gonoskov, A. and Blackburn, T. G. and Marklund, M. and Bulanov, S. S.},
  journal = {Rev. Mod. Phys.},
  volume = {94},
  issue = {4},
  pages = {045001},
  numpages = {63},
  year = {2022},
  month = {Oct},
  publisher = {American Physical Society},
  doi = {10.1103/RevModPhys.94.045001},
  url = {https://link.aps.org/doi/10.1103/RevModPhys.94.045001}
}

@inbook{schwinger1945,
  author={Schwinger, J.},
  year= 2000, 
  chapter={On radiation by electrons in a betatron, {R}eport {N}o. {LBNL}--39088}, 
  editor = {Kimball A. Milton}, 
  title= {A {Q}uantum {L}egacy: {S}eminal {P}apers of {J}ulian {S}chwinger}, 
  publisher= {World Scientific},
  address= {Singapore}, 
}

@article{michel_prl1982,
  title = {Intense Coherent Submillimeter Radiation in Electron Storage Rings},
  author = {Michel, F. Curtis},
  journal = {Phys. Rev. Lett.},
  volume = {48},
  issue = {9},
  pages = {580--583},
  numpages = {0},
  year = {1982},
  month = {Mar},
  publisher = {American Physical Society},
  doi = {10.1103/PhysRevLett.48.580},
  url = {https://link.aps.org/doi/10.1103/PhysRevLett.48.580}
}

@article{hirschmugl_pra1991,
  title = {Multiparticle coherence calculations for synchrotron-radiation emission},
  author = {Hirschmugl, Carol J. and Sagurton, Michael and Williams, Gwyn P.},
  journal = {Phys. Rev. A},
  volume = {44},
  issue = {2},
  pages = {1316--1320},
  numpages = {0},
  year = {1991},
  month = {Jul},
  publisher = {American Physical Society},
  doi = {10.1103/PhysRevA.44.1316},
  url = {https://link.aps.org/doi/10.1103/PhysRevA.44.1316}
}

@article{gonoskov_prl2014,
  title = {Anomalous Radiative Trapping in Laser Fields of Extreme Intensity},
  author = {Gonoskov, A. and Bashinov, A. and Gonoskov, I. and Harvey, C. and Ilderton, A. and Kim, A. and Marklund, M. and Mourou, G. and Sergeev, A.},
  journal = {Phys. Rev. Lett.},
  volume = {113},
  issue = {1},
  pages = {014801},
  numpages = {5},
  year = {2014},
  month = {Jul},
  publisher = {American Physical Society},
  doi = {10.1103/PhysRevLett.113.014801},
  url = {https://link.aps.org/doi/10.1103/PhysRevLett.113.014801}
}

@article{ji_prl2014,
  title = {Radiation-Reaction Trapping of Electrons in Extreme Laser Fields},
  author = {Ji, L. L. and Pukhov, A. and Kostyukov, I. Yu. and Shen, B. F. and Akli, K.},
  journal = {Phys. Rev. Lett.},
  volume = {112},
  issue = {14},
  pages = {145003},
  numpages = {5},
  year = {2014},
  month = {Apr},
  publisher = {American Physical Society},
  doi = {10.1103/PhysRevLett.112.145003},
  url = {https://link.aps.org/doi/10.1103/PhysRevLett.112.145003}
}

@article{chen_ppcf2010,
  title={Radiation reaction effects on ion acceleration in laser foil interaction},
  author={Chen, Min and Pukhov, Alexander and Yu, Tong-Pu and Sheng, Zheng-Ming},
  journal={Plasma Physics and Controlled Fusion},
  volume={53},
  number={1},
  pages={014004},
  year={2010},
  publisher={IOP Publishing}
}

@article{gelfer_scirep2018,
  title={Unexpected impact of radiation friction: enhancing production of longitudinal plasma waves},
  author={Gelfer, Evgeny and Elkina, Nina and Fedotov, Alexander},
  journal={Sci. Rep.},
  volume={8},
  number={1},
  pages={6478},
  year={2018},
  publisher={Nature Publishing Group}
}

@article{gelfer_ppcf2018,
  title={Theory and simulations of radiation friction induced enhancement of laser-driven longitudinal fields},
  author={Gelfer, EG and Fedotov, AM and Weber, S},
  journal={Plasma Phys. Controlled Fusion},
  volume={60},
  number={6},
  pages={064005},
  year={2018},
  publisher={IOP Publishing}
}

@article{liseykina_njp2016,
  title={Inverse {F}araday effect driven by radiation friction},
  author={Liseykina, TV and Popruzhenko, SV and Macchi, A},
  journal={New J. Phys.},
  volume={18},
  number={7},
  pages={072001},
  year={2016},
  publisher={IOP Publishing}
}

@article{gelfer_njp2021,
  title={Radiation induced acceleration of ions in a laser irradiated transparent foil},
  author={Gelfer, EG and Fedotov, AM and Weber, S},
  journal={New J. Phys.},
  volume={23},
  number={9},
  pages={095002},
  year={2021},
  publisher={IOP Publishing}
}

@article{esarey_pre1993,
  title = {Nonlinear {Thomson} scattering of intense laser pulses from beams and plasmas},
  author = {Esarey, Eric and Ride, Sally K. and Sprangle, Phillip},
  journal = {Phys. Rev. E},
  volume = {48},
  issue = {4},
  pages = {3003--3021},
  numpages = {0},
  year = {1993},
  month = {Oct},
  publisher = {American Physical Society},
  doi = {10.1103/PhysRevE.48.3003},
  url = {https://link.aps.org/doi/10.1103/PhysRevE.48.3003}
}

@article{dirac_prs1938,
  title={Classical theory of radiating electrons},
  author={Dirac, Paul Adrien Maurice},
  journal={Proc. R. Soc. A},
  volume={167},
  number={929},
  pages={148--169},
  year={1938},
  publisher={The Royal Society London}
}

@book{jackson_book,
  title={Classical electrodynamics},
  author={Jackson, John David},
  year={2021},
  publisher={John Wiley \& Sons}
}

@article{burton_contphys2014,
  title={Aspects of electromagnetic radiation reaction in strong fields},
  author={Burton, David A and Noble, Adam},
  journal={Contemp. Phys.},
  volume={55},
  number={2},
  pages={110--121},
  year={2014},
  publisher={Taylor \& Francis}
}

@article{koga_pop2005,
    author = {Koga, James and Esirkepov, Timur Zh. and Bulanov, Sergei V.},
    title = "{Nonlinear {T}homson scattering in the strong radiation damping regime}",
    journal = {Physics of Plasmas},
    volume = {12},
    number = {9},
    pages = {093106},
    year = {2005},
    month = {09},
    issn = {1070-664X},
    doi = {10.1063/1.2013067},
    url = {https://doi.org/10.1063/1.2013067},
}

@article{dipiazza_lmp2008,
  title={Exact solution of the {L}andau--{L}ifshitz equation in a plane wave},
  author={Piazza, A Di},
  journal={Lett. Math. Phys.},
  volume={83},
  pages={305--313},
  year={2008},
  publisher={Springer}
}

@article{voronin_jetp1965,
  title={The pressure of an intense plane wave on a free charge and on a charge in a magnetic field},
  author={Voronin, BS and Kolomenskii, AA},
  journal={Sov. Phys. JETP},
  volume={20},
  pages={1027},
  year={1965}
}

@article{zeldovich_ufn1975,
  title={Interaction of free electrons with electromagnetic radiation},
  author={Zel'Dovich, Ya B},
  journal={Sov. Phys. Usp.},
  volume={18},
  number={2},
  pages={79},
  year={1975},
  publisher={IOP Publishing}
}

@article{popruzhenko_njp2019,
  title={Efficiency of radiation friction losses in laser-driven ‘hole boring’of dense targets},
  author={Popruzhenko, SV and Liseykina, TV and Macchi, A},
  journal={New J. Phys.},
  volume={21},
  number={3},
  pages={033009},
  year={2019},
  publisher={IOP Publishing}
}

@article{formanek_pra2022,
  title = {Radiation reaction enhancement in flying focus pulses},
  author = {Formanek, M. and Ramsey, D. and Palastro, J. P. and Di Piazza, A.},
  journal = {Phys. Rev. A},
  volume = {105},
  issue = {2},
  pages = {L020203},
  numpages = {6},
  year = {2022},
  month = {Feb},
  publisher = {American Physical Society},
  doi = {10.1103/PhysRevA.105.L020203},
  url = {https://link.aps.org/doi/10.1103/PhysRevA.105.L020203}
}

@article{gelfer_prr2024,
  title = {Coherent radiation of an electron bunch colliding with an intense laser pulse},
  author = {Gelfer, E. G. and Fedotov, A. M. and Klimo, O. and Weber, S.},
  journal = {Phys. Rev. Res.},
  volume = {6},
  issue = {3},
  pages = {L032013},
  numpages = {7},
  year = {2024},
  month = {Jul},
  publisher = {American Physical Society},
  doi = {10.1103/PhysRevResearch.6.L032013},
  url = {https://link.aps.org/doi/10.1103/PhysRevResearch.6.L032013}
}

@article{gelfer_mre2024,
    author = {Gelfer, E. G. and Fedotov, A. M. and Klimo, O. and Weber, S.},
    title = "{Collective coherent emission of electrons in strong laser fields and perspective for hard x-ray lasers}",
    journal = {Matter Radiat. Extremes},
    volume = {9},
    number = {2},
    pages = {024201},
    year = {2024},
    month = {02},
    issn = {2468-2047},
    doi = {10.1063/5.0174508},
    url = {https://doi.org/10.1063/5.0174508},
}

@article{hartemann_pre2000,
  title={Stochastic electron gas theory of coherence in laser-driven synchrotron radiation},
  author={Hartemann, FV},
  journal={Phys. Rev. E},
  volume={61},
  number={1},
  pages={972},
  year={2000},
  publisher={APS}
}

@article{schiff_rsi1946,
    author = {Schiff, L. I.},
    title = "{Production of Particle Energies beyond 200 {Mev}}",
    journal = {Rev. Sci. Instrum.},
    volume = {17},
    number = {1},
    pages = {6-14},
    year = {1946},
    month = {01},
    issn = {0034-6748},
    doi = {10.1063/1.1770395},
    url = {https://doi.org/10.1063/1.1770395},
}

@article{thomas_prstab2010,
  title = {Algorithm for calculating spectral intensity due to charged particles in arbitrary motion},
  author = {Thomas, A. G. R.},
  journal = {Phys. Rev. ST Accel. Beams},
  volume = {13},
  issue = {2},
  pages = {020702},
  numpages = {11},
  year = {2010},
  month = {Feb},
  publisher = {American Physical Society},
  doi = {10.1103/PhysRevSTAB.13.020702},
  url = {https://link.aps.org/doi/10.1103/PhysRevSTAB.13.020702}
}

@article{kharin_pra2016,
  title = {Temporal laser-pulse-shape effects in nonlinear {T}homson scattering},
  author = {Kharin, V. Yu. and Seipt, D. and Rykovanov, S. G.},
  journal = {Phys. Rev. A},
  volume = {93},
  issue = {6},
  pages = {063801},
  numpages = {8},
  year = {2016},
  month = {Jun},
  publisher = {American Physical Society},
  doi = {10.1103/PhysRevA.93.063801},
  url = {https://link.aps.org/doi/10.1103/PhysRevA.93.063801}
}

@article{seipt_lasphys2013,
doi = {10.1088/1054-660X/23/7/075301},
url = {https://dx.doi.org/10.1088/1054-660X/23/7/075301},
year = {2013},
month = {may},
publisher = {IOP Publishing},
volume = {23},
number = {7},
pages = {075301},
author = {D Seipt and B Kämpfer},
title = {Nonlinear {C}ompton scattering of ultrahigh-intensity laser pulses},
journal = {Laser Phys.},
}

@article{boca_pra2009,
  title = {Nonlinear {Compton} scattering with a laser pulse},
  author = {Boca, Madalina and Florescu, Viorica},
  journal = {Phys. Rev. A},
  volume = {80},
  issue = {5},
  pages = {053403},
  numpages = {14},
  year = {2009},
  month = {Nov},
  publisher = {American Physical Society},
  doi = {10.1103/PhysRevA.80.053403},
  url = {https://link.aps.org/doi/10.1103/PhysRevA.80.053403}
}

@article{seipt_pra2011,
  title = {Nonlinear {Compton} scattering of ultrashort intense laser pulses},
  author = {Seipt, D. and K\"ampfer, B.},
  journal = {Phys. Rev. A},
  volume = {83},
  issue = {2},
  pages = {022101},
  numpages = {12},
  year = {2011},
  month = {Feb},
  publisher = {American Physical Society},
  doi = {10.1103/PhysRevA.83.022101},
  url = {https://link.aps.org/doi/10.1103/PhysRevA.83.022101}
}

@article{gelfer_prd2022,
  title = {Nonlinear {Compton} scattering in time-dependent electric fields beyond the locally constant crossed field approximation},
  author = {Gelfer, E. G. and Fedotov, A. M. and Mironov, A. A. and Weber, S.},
  journal = {Phys. Rev. D},
  volume = {106},
  issue = {5},
  pages = {056013},
  numpages = {16},
  year = {2022},
  month = {Sep},
  publisher = {American Physical Society},
  doi = {10.1103/PhysRevD.106.056013},
  url = {https://link.aps.org/doi/10.1103/PhysRevD.106.056013}
}

@article{salehi_prx2021,
  title = {Laser-Accelerated, Low-Divergence 15-{MeV} Quasimonoenergetic Electron Bunches at 1 {kHz}},
  author = {Salehi, F. and Le, M. and Railing, L. and Kolesik, M. and Milchberg, H. M.},
  journal = {Phys. Rev. X},
  volume = {11},
  issue = {2},
  pages = {021055},
  numpages = {12},
  year = {2021},
  month = {Jun},
  publisher = {American Physical Society},
  doi = {10.1103/PhysRevX.11.021055},
  url = {https://link.aps.org/doi/10.1103/PhysRevX.11.021055}
}

@article{storey_prstab2024,
  title = {Wakefield generation in hydrogen and lithium plasmas at {FACET-II}: Diagnostics and first beam-plasma interaction results},
  author = {Storey, D. and Zhang, C. and San Miguel Claveria, P. and Cao, G. J. and Adli, E. and Alsberg, L. and Ariniello, R. and Clarke, C. and Corde, S. and Dalichaouch, T. N. and others},
  journal = {Phys. Rev. Accel. Beams},
  volume = {27},
  issue = {5},
  pages = {051302},
  numpages = {15},
  year = {2024},
  month = {May},
  publisher = {American Physical Society},
  doi = {10.1103/PhysRevAccelBeams.27.051302},
  url = {https://link.aps.org/doi/10.1103/PhysRevAccelBeams.27.051302}
}

@article{yakimenko_prl2019,
  title = {Prospect of Studying Nonperturbative {QED} with Beam-Beam Collisions},
  author = {Yakimenko, V. and Meuren, S. and Del Gaudio, F. and Baumann, C. and Fedotov, A. and Fiuza, F. and Grismayer, T. and Hogan, M. J. and Pukhov, A. and Silva, L. O. and White, G.},
  journal = {Phys. Rev. Lett.},
  volume = {122},
  issue = {19},
  pages = {190404},
  numpages = {7},
  year = {2019},
  month = {May},
  publisher = {American Physical Society},
  doi = {10.1103/PhysRevLett.122.190404},
  url = {https://link.aps.org/doi/10.1103/PhysRevLett.122.190404}
}

@article{chang_prappl2023,
  title = {Reduction of the electron-beam divergence of laser wakefield accelerators by integrated plasma lenses},
  author = {Chang, Y.-Y. and Cabada\ifmmode \breve{g}\else \u{g}\fi{}, J. Couperus and Debus, A. and Ghaith, A. and LaBerge, M. and Pausch, R. and Sch\"obel, S. and Ufer, P. and Schramm, U. and Irman, A.},
  journal = {Phys. Rev. Appl.},
  volume = {20},
  issue = {6},
  pages = {L061001},
  numpages = {6},
  year = {2023},
  month = {Dec},
  publisher = {American Physical Society},
  doi = {10.1103/PhysRevApplied.20.L061001},
  url = {https://link.aps.org/doi/10.1103/PhysRevApplied.20.L061001}
}

@article{sarri_ncomms2015,
  title={Generation of neutral and high-density electron--positron pair plasmas in the laboratory},
  author={Sarri, Gianluca and Poder, K and Cole, JM and Schumaker, W and Di Piazza, Antonino and Reville, Brian and Dzelzainis, T and Doria, D and Gizzi, LA and Grittani, G and others},
  journal={Nat. Commun.},
  volume={6},
  number={1},
  pages={6747},
  year={2015},
  publisher={Nature Publishing Group UK London}
}

@online{facet2,
    howpublished = {\url{https://facet-ii.slac.stanford.edu/facility/beam-parameters}},
    title = {{FACET II} beam parameters}
}

@article{polyanskiy_osacont2020,
  title={Demonstration of a 2 ps, 5 {TW} peak power, long-wave infrared laser based on chirped-pulse amplification with mixed-isotope $CO_2$ amplifiers},
  author={Polyanskiy, Mikhail N and Pogorelsky, Igor V and Babzien, Marcus and Palmer, Mark A},
  journal={OSA Continuum},
  volume={3},
  number={3},
  pages={459--472},
  year={2020},
  publisher={Optica Publishing Group}
}

@article{haberberger_optexpr2010,
  title={Fifteen terawatt picosecond $CO_2$ laser system},
  author={Haberberger, D and Tochitsky, S and Joshi, C},
  journal={Opt. Express},
  volume={18},
  number={17},
  pages={17865--17875},
  year={2010},
  publisher={Optica Publishing Group}
}

@article{chang_advoptphot2022,
  title={Intense infrared lasers for strong-field science},
  author={Chang, Zenghu and Fang, Li and Fedorov, Vladimir and Geiger, Chase and Ghimire, Shambhu and Heide, Christian and Ishii, Nobuhisa and Itatani, Jiro and Joshi, Chandrashekhar and Kobayashi, Yuki and others},
  journal={Adv. Opt. Photonics},
  volume={14},
  number={4},
  pages={652--782},
  year={2022},
  publisher={Optica Publishing Group}
}

@phdthesis{quin_phd2023,
  author  = "Quin, M J",
  title   = "Classical Radiation Reaction and Collective Behaviour",
  school  = "University of Heidelberg",
  year    = "2023",
  url="https://archiv.ub.uni-heidelberg.de/volltextserver/34028/"
}

@article{woodward_jiee1946,
  title={A method of calculating the field over a plane aperture required to produce a given polar diagram},
  author={Woodward, PM},
  journal={J. Inst. Electr. Eng. -- Part IIIA: Radiolocation},
  volume={93},
  number={10},
  pages={1554--1558},
  year={1946},
  publisher={IET}
}

@article{lawson_ieee1979,
  title={Lasers and accelerators},
  author={Lawson, John David},
  journal={IEEE Trans. Nucl. Sci.},
  volume={26},
  number={3},
  pages={4217--4219},
  year={1979},
  publisher={IEEE}
}

@article{kirk_ppcf2009,
  title={Pair production in counter-propagating laser beams},
  author={Kirk, John G and Bell, AR and Arka, Ioanna},
  journal={Plasma Physics and Controlled Fusion},
  volume={51},
  number={8},
  pages={085008},
  year={2009},
  publisher={IOP Publishing}
}

@article{tamburini_nimpra2011,
  title={Radiation reaction effects on electron nonlinear dynamics and ion acceleration in laser--solid interaction},
  author={Tamburini, Matteo and Pegoraro, Francesco and Di Piazza, Antonino and Keitel, Christoph H and Liseykina, Tatyana V and Macchi, Andrea},
  journal={Nucl. Instrum. Methods Phys. Res., Sect. A},
  volume={653},
  number={1},
  pages={181--185},
  year={2011},
  publisher={Elsevier}
}

@article{neitz_prl2013,
  title={Stochasticity effects in quantum radiation reaction},
  author={Neitz, Norman and Di Piazza, Antonino},
  journal={Phys. Rev. Lett.},
  volume={111},
  number={5},
  pages={054802},
  year={2013},
  publisher={APS}
}

@article{tamburini_pre2012,
  title={Radiation-pressure-dominant acceleration: Polarization and radiation reaction effects and energy increase in three-dimensional simulations},
  author={Tamburini, M and Liseykina, TV and Pegoraro, Francesco and Macchi, A},
  journal={Phys. Rev. E},
  volume={85},
  number={1},
  pages={016407},
  year={2012},
  publisher={APS}
}

@article{tamburini_njp2010,
  title={Radiation reaction effects on radiation pressure acceleration},
  author={Tamburini, M and Pegoraro, Francesco and Di Piazza, A and Keitel, Ch H and Macchi, A},
  journal={New J. Phys.},
  volume={12},
  number={12},
  pages={123005},
  year={2010},
  publisher={IOP Publishing}
}

@article{capdessus_pre2015,
  title={Influence of radiation reaction force on ultraintense laser-driven ion acceleration},
  author={Capdessus, R and McKenna, Paul},
  journal={Phys. Rev. E},
  volume={91},
  number={5},
  pages={053105},
  year={2015},
  publisher={APS}
}

@article{vranic_njp2016,
  title={Quantum radiation reaction in head-on laser-electron beam interaction},
  author={Vranic, Marija and Grismayer, Thomas and Fonseca, Ricardo A and Silva, Luis O},
  journal={New J. Phys.},
  volume={18},
  number={7},
  pages={073035},
  year={2016},
  publisher={IOP Publishing}
}

@article{dipiazza_prl2010,
  title={Quantum radiation reaction effects in multiphoton {Compton} scattering},
  author={Di Piazza, A and Hatsagortsyan, KZ and Keitel, Christoph H},
  journal={Phys. Rev. Lett.},
  volume={105},
  number={22},
  pages={220403},
  year={2010},
  publisher={APS}
}

@article{thomas_prx2012,
  title={Strong radiation-damping effects in a gamma-ray source generated by the interaction of a high-intensity laser with a wakefield-accelerated electron beam},
  author={Thomas, AGR and Ridgers, CP and Bulanov, SS and Griffin, BJ and Mangles, SPD},
  journal={Phys. Rev. X},
  volume={2},
  number={4},
  pages={041004},
  year={2012},
  publisher={APS}
}

@article{shen_prl1972,
  title = {Energy Straggling and Radiation Reaction for Magnetic Bremsstrahlung},
  author = {Shen, C. S. and White, D.},
  journal = {Phys. Rev. Lett.},
  volume = {28},
  issue = {7},
  pages = {455--459},
  numpages = {0},
  year = {1972},
  month = {Feb},
  publisher = {American Physical Society},
  doi = {10.1103/PhysRevLett.28.455},
  url = {https://link.aps.org/doi/10.1103/PhysRevLett.28.455}
}

@article{blackburn_prl2014,
  title = {Quantum Radiation Reaction in Laser--Electron-Beam Collisions},
  author = {Blackburn, T. G. and Ridgers, C. P. and Kirk, J. G. and Bell, A. R.},
  journal = {Phys. Rev. Lett.},
  volume = {112},
  issue = {1},
  pages = {015001},
  numpages = {5},
  year = {2014},
  month = {Jan},
  publisher = {American Physical Society},
  doi = {10.1103/PhysRevLett.112.015001},
  url = {https://link.aps.org/doi/10.1103/PhysRevLett.112.015001}
}

@article{harvey_prl2017,
  title = {Quantum Quenching of Radiation Losses in Short Laser Pulses},
  author = {Harvey, C. N. and Gonoskov, A. and Ilderton, A. and Marklund, M.},
  journal = {Phys. Rev. Lett.},
  volume = {118},
  issue = {10},
  pages = {105004},
  numpages = {5},
  year = {2017},
  month = {Mar},
  publisher = {American Physical Society},
  doi = {10.1103/PhysRevLett.118.105004},
  url = {https://link.aps.org/doi/10.1103/PhysRevLett.118.105004}
}

@article{vieira_natphys2021,
  title={Generalized superradiance for producing broadband coherent radiation with transversely modulated arbitrarily diluted bunches},
  author={Vieira, J and Pardal, Miguel and Mendon{\c{c}}a, JT and Fonseca, RA},
  journal={Nat. Phys.},
  volume={17},
  number={1},
  pages={99--104},
  year={2021},
  publisher={Nature Publishing Group UK London}
}

@book{birdsall_book2018,
  title={Plasma physics via computer simulation},
  author={Birdsall, Charles K and Langdon, A Bruce},
  year={2018},
  publisher={CRC press}
}

@article{dawson_rmp1983,
  title={Particle simulation of plasmas},
  author={Dawson, John M},
  journal={Rev. Mod. Phys.},
  volume={55},
  number={2},
  pages={403},
  year={1983},
  publisher={APS}
}

@article{quin_ppcf2025,
  title={Coherent frequency combs from electrons colliding with a laser pulse},
  author={Quin, Michael J and Di Piazza, Antonino and Tamburini, Matteo},
  journal={Plasma Physics and Controlled Fusion},
  volume={67},
  number={5},
  pages={055008},
  year={2025},
  publisher={IOP Publishing}
}

@article{quin_prr2025,
  title = {Effect of interparticle fields and radiation reaction on beam dynamics},
  author = {Quin, Michael J. and Di Piazza, Antonino and Keitel, Christoph H. and Tamburini, Matteo},
  journal = {Phys. Rev. Res.},
  volume = {7},
  issue = {2},
  pages = {023210},
  numpages = {12},
  year = {2025},
  month = {Jun},
  publisher = {American Physical Society},
  doi = {10.1103/PhysRevResearch.7.023210},
  url = {https://link.aps.org/doi/10.1103/PhysRevResearch.7.023210}
}

@article{malaca_natphot2024,
  title={Coherence and superradiance from a plasma-based quasiparticle accelerator},
  author={Malaca, B and Pardal, M and Ramsey, D and Pierce, JR and Weichman, K and Andriyash, IA and Mori, WB and Palastro, JP and Fonseca, RA and Vieira, J},
  journal={Nat. Phot.},
  volume={18},
  number={1},
  pages={39--45},
  year={2024},
  publisher={Nature Publishing Group UK London},
  url={https://www.nature.com/articles/s41566-023-01311-z#citeas}
}

@article{sakai_scirep2024,
  title={Hard {X}-ray inverse {C}ompton scattering at photon energy of 87.5 ke{V}},
  author={Sakai, Yusuke and Babzien, Marcus and Fedurin, Mikhail and Kusche, Karl and Williams, Oliver and Fukasawa, Atsushi and Naranjo, Brian and Murokh, Alex and Agustsson, Ronald and Simmonds, Andrew and Jacob, Paul and Stenby, George and  Malone, Robert and  Polyanskiy, Mikhail and Pogorelsky, Igor and  Palmer, Mark and  Rosenzweig, James},
  journal={Scientific Reports},
  volume={14},
  number={1},
  pages={18467},
  year={2024},
  doi={https://doi.org/10.1038/s41598-024-68170-8},
  publisher={Nature Publishing Group UK London}
}

@article{seipt_prl2017,
  title={Depletion of intense fields},
  author={Seipt, D and Heinzl, T and Marklund, M and Bulanov, SS},
  journal={Physical Review Letters},
  volume={118},
  number={15},
  pages={154803},
  year={2017},
  publisher={APS}
}

@article{pre_arxiv,
  title={Analytical theory of coherent radiation and radiation friction in laser--plasma collisions},
  author={Gelfer, EG and Fedotov, AM and Malakhov, MP and Benahmed, Th and Custodio, J and Klimo, O and Weber, S.},
  journal={arXiv preprint arXiv:2605.28370},
  year={2026},
}

@article{los_natcomm2026,
  title={Observation of quantum effects on radiation reaction in strong fields},
  author={Los, Eva E and Gerstmayr, Elias and Arran, Christopher and Streeter, Matthew JV and Colgan, Cary and Cobo, Claudia C and Kettle, Brendan and Blackburn, Thomas G and Bourgeois, Nicolas and Calvin, Luke and others},
  journal={Nature Communications},
  year={2026},
  publisher={Nature Publishing Group UK London}
}

@article{los_hplse2025,
  title={A Bayesian framework to investigate radiation reaction in strong fields},
  author={Los, Eva E and Arran, Christopher and Gerstmayr, Elias and Streeter, Matthew JV and Kettle, Brendan and Najmudin, Zulfikar and Ridgers, Christopher P and Sarri, Gianluca and Mangles, Stuart PD},
  journal={High Power Laser Science and Engineering},
  volume={13},
  pages={e25},
  year={2025},
  publisher={Cambridge University Press}
}

@article{blackman_commphys2022,
  title={Electron acceleration from transparent targets irradiated by ultra-intense helical laser beams},
  author={Blackman, David R and Shi, Yin and Klein, Sallee R and Cernaianu, Mihail and Doria, Domenico and Ghenuche, Petru and Arefiev, Alexey},
  journal={Communications Physics},
  volume={5},
  number={1},
  pages={116},
  year={2022},
  publisher={Nature Publishing Group UK London}
}

@article{shi_hplse2022,
  title={Electron pulse train accelerated by a linearly polarized Laguerre--Gaussian laser beam},
  author={Shi, Yin and Blackman, David R and Zhu, Ping and Arefiev, Alexey},
  journal={High Power Laser Science and Engineering},
  volume={10},
  pages={e45},
  year={2022},
  publisher={Cambridge University Press}
}

@article{shi_ppcf2021,
  title={Electron acceleration using twisted laser wavefronts},
  author={Shi, Yin and R Blackman, David and Arefiev, Alexey},
  journal={Plasma Physics and Controlled Fusion},
  volume={63},
  number={12},
  pages={125032},
  year={2021},
  publisher={IOP Publishing}
}

@article{shi_prl2021,
  title={Generation of ultrarelativistic monoenergetic electron bunches via a synergistic interaction of longitudinal electric and magnetic fields of a twisted laser},
  author={Shi, Yin and Blackman, David and Stutman, Dan and Arefiev, Alexey},
  journal={Physical Review Letters},
  volume={126},
  number={23},
  pages={234801},
  year={2021},
  publisher={APS}
}

\end{document}